\documentclass[11pt]{article}
\usepackage[final]{acl}

\title{Can Large Audio-Language Models Ignore Multilingual Distractors? \\ An Evaluation of Their Selective Auditory Attention Capabilities}

\author{Heejoon Koo \\
  University of Illinois Urbana-Champaign \\
  \texttt{heejoon4@illinois.edu} \\}

\usepackage{comment}
\usepackage{multirow}
\usepackage{makecell}
\usepackage{graphicx}
\usepackage{booktabs}
\usepackage{times}
\usepackage{latexsym}
\usepackage[T1]{fontenc}
\usepackage[utf8]{inputenc}
\usepackage{microtype}
\usepackage{inconsolata}
\usepackage{graphicx}
\usepackage{enumitem}
\usepackage[most]{tcolorbox}
\usepackage{xcolor}
\usepackage{colortbl}
\usepackage[table]{xcolor}
\usepackage{CJKutf8}

\newtcolorbox{examplebox}[1][]{
  enhanced,
  breakable,
  colback=gray!4,
  colframe=black!55,
  boxrule=0.5pt,
  arc=2pt,
  left=6pt,
  right=6pt,
  top=6pt,
  bottom=6pt,
  title=#1,
  fonttitle=\bfseries,
  coltitle=black,
  colbacktitle=gray!18
}

\newtcolorbox{promptbox}[1][]{
  enhanced,
  breakable,
  colback=gray!4,
  colframe=black!55,
  boxrule=0.5pt,
  arc=2pt,
  left=6pt,
  right=6pt,
  top=6pt,
  bottom=6pt,
  title=#1,
  fonttitle=\bfseries,
  coltitle=black,
  colbacktitle=gray!18
}

\usepackage{fvextra}

\DefineVerbatimEnvironment{PromptVerbatim}{Verbatim}{
    breaklines=true,
    breaksymbolleft={},
    fontsize=\small,
    frame=single,
    rulecolor=\color{gray},
    framesep=6pt
}

\begin{document}

\maketitle

\begin{abstract}
    Robust selective auditory attention under multilingual interference is critical for reliable LALM deployment. We introduce MUSI, a cocktail party-inspired controlled diagnostic evaluation of source-grounded spoken-language understanding and reasoning. Each item pairs an English target dialogue with a plausible distractor in English, Spanish, Korean, or Chinese, and evaluates models under (1) single-stream, (2) separation-based, and (3) end-to-end cocktail party settings across controlled SNRs. Across four open-weight and two closed-source LALMs, we find model-dependent language-conditioned variation and heightened vulnerability at adverse SNRs. Errors are dominated by distractor-grounded source confusion, while separation reduces acoustic overlap but often leaves source attribution unresolved. These findings highlight selective auditory attention as a primary capability for reliable LALM deployment.
\end{abstract}


\section{Introduction}

Overlapping speech has long been studied in ASR and speech separation through multi-speaker recognition, diarization, and source separation \cite{hershey2016deep, cosentino2020librimix, watanabe2020chime}. Auditory neuroscience frames the same challenge as the cocktail party problem: listeners must organize acoustic mixtures and selectively attend to task-relevant sources while suppressing competing speech \cite{cherry1953some, mcdermott2009cocktail, bronkhorst2015cocktail}. For Large Audio-Language Models (LALMs), however, separating or transcribing speech does not guarantee source-grounded reasoning: a model may rely on the wrong stream, conflate sources, or generate fluent but source-incoherent answers.

\begin{figure}[t]
    \centering
    \includegraphics[width=0.85\linewidth]{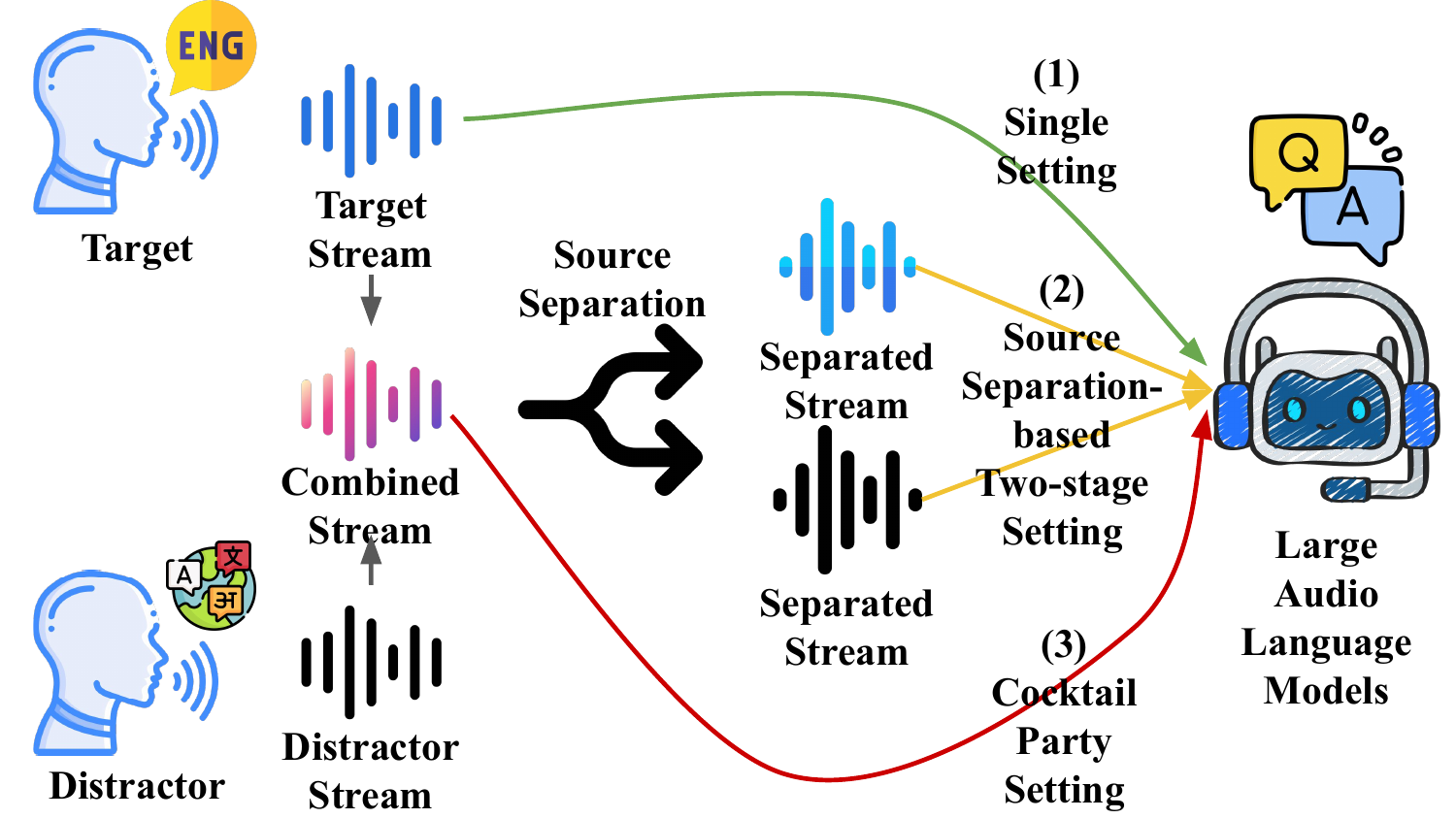}
    \caption{Our MUSI evaluation framework.}
    \label{fig:example}
\vspace{-1em}
\end{figure}

This failure mode is critical in high-stakes domains such as aviation and healthcare, where audio-based systems are increasingly used for tasks including respiratory sound analysis \cite{koo2026mitigating, koo2026empowering}, and concurrent same- or cross-lingual speech may contain semantically salient but task-irrelevant information.
It also raises a multilingual reliability concern: performance may vary with the language of surrounding speech. Existing benchmarks address related but distinct capabilities: multi-speaker ASR and separation benchmarks focus on transcription, diarization, or signal reconstruction \cite{hershey2016deep, watanabe2020chime, cosentino2020librimix, borsdorf2021target, nguyen2026cocktail}; general LALM benchmarks primarily evaluate single-stream audio understanding \cite{yang2024air, wang2025audiobench, yang2025sakura, chen2026voicebench}; and trustworthiness benchmarks study acoustic perturbations or instruction-level threats \cite{li2025audiotrust, song2025audio, hou2025evaluating, cheng2025jailbreak, peng2025jalmbench}. As summarized in Table~\ref{tab:related}, none directly evaluates whether LALMs can follow a target dialogue while suppressing plausible multilingual distractors. Such controlled interference reveals whether answers are grounded in the target, a competing stream, or neither, and how performance varies across distractor languages.

\begin{table*}[t]
\centering
\scriptsize
\setlength{\tabcolsep}{5pt}
\renewcommand{\arraystretch}{1.15}
\begin{tabular}{p{2.3cm}p{2.4cm}p{4.2cm}p{5.0cm}}
\toprule
\midrule
\textbf{Category} & \textbf{Literature} & \textbf{Task} & \textbf{Gap relative to MUSI} \\
\hline
\midrule
\multirow{2}{=}{Multi-speaker ASR \& Separation}
  & CHiME-6 \cite{watanabe2020chime}
  & Multi-speaker far-field ASR and diarization
  & Transcription/diarization quality, not source-grounded reasoning \\
  & WSJ0-2mix \cite{hershey2016deep} / LibriMix \cite{cosentino2020librimix}
  & Single-channel two-speaker separation
  & Signal reconstruction (SI-SDR); no LALM reasoning \\
\midrule
\multirow{2}{=}{Cocktail Party Listening}
  & TLE \cite{borsdorf2021target}
  & Target-language extraction in multilingual mixtures
  & Language-conditioned separation; no MCQ reasoning or source attribution \\
  & MCoRec \cite{nguyen2026cocktail}
  & Multi-modal multi-party conversation transcription
  & Transcription- and clustering-oriented; no reasoning or source attribution \\
\midrule
\multirow{3}{=}{General LALM Evaluation}
  & AIR-Bench \cite{yang2024air}
  & LALM audio understanding (MCQ + chat)
  & Clean single-stream audio; no concurrent distractors \\
  & AudioBench \cite{wang2025audiobench}
  & Universal LALM benchmark, 8 tasks
  & Single-source audio; no multilingual interference \\
  & VoiceBench \cite{chen2026voicebench}
  & Voice assistant evaluation under varied conditions
  & Acoustic/speaker variations but no semantic concurrent distractors \\
\midrule
\multirow{2}{=}{Trustworthiness \& Reasoning}
  & AudioTrust \cite{li2025audiotrust}
  & LALM trustworthiness across 6 dimensions
  & Non-semantic acoustic risks; not source misattribution \\
  & SAKURA \cite{yang2025sakura}
  & Multi-hop reasoning over audio attributes
  & Single-stream input; no selective attention under interference \\
\midrule
\rowcolor{gray!10}
\textbf{Source-grounded QA} & \textbf{MUSI (Ours)}
  & \textbf{MCQ-based selective attention with concurrent multilingual semantic distractors}
  & \textbf{---} \\
\bottomrule
\end{tabular}
\caption{Positioning of MUSI across related work, from signal-level multi-speaker ASR and source separation to recent LALM benchmarks.}
\label{tab:related}
\vspace{-1em}
\end{table*}

We introduce \textbf{MUSI} (\textbf{Mu}ltilingual \textbf{S}elective Attent\textbf{I}on), a multilingual multiple-choice question-answering (MCQ) testbed for source-grounded LALM understanding and reasoning. Each item pairs an English target dialogue with a concurrent distractor in English, Spanish, Korean, or Chinese, covering both same-language and cross-lingual interference \cite{chiswick2005linguistic, littell2017uriel}. We evaluate six LALMs under single-stream, separation-based two-stage, and end-to-end cocktail party settings, with controlled target-to-distractor SNRs.

Our results reveal substantial model-dependent variation across distractor
languages and strong sensitivity to adverse SNRs. Errors are dominated by distractor-grounded source confusion, while confidence becomes miscalibrated under interference for open-weight models. Although separation reduces acoustic overlap, it often leaves source attribution unresolved and yields confident wrong-stream answers.

In summary, our contributions are as follows:
\begin{itemize}[leftmargin=*, itemsep=1pt, topsep=2pt, parsep=0pt]
\item We introduce MUSI, a multilingual controlled diagnostic evaluation for target-grounded reasoning under plausible concurrent speech distractors.
\item We characterize model-dependent variation across distractor languages and trace model errors to the target, distractor, or neither stream.
\item We show that strong single-stream performance does not ensure cocktail party robustness and that separation-based inference retains a source-attribution bottleneck.
\end{itemize}


\section{MUSI Dataset}

\subsection{Dataset Composition}

MUSI covers four safety-critical domains: Aviation \& Maritime, Healthcare \& Medicine, Finance \& Business, and Construction \& Plant. Each item contains an English target dialogue, a same-domain semantically plausible distractor in English, Spanish, Korean, or Chinese, a target-grounded question, and four answer options. The question is answerable only from the target, whereas the distractor provides plausible but task-irrelevant evidence. MUSI contains 200 cases, with 50 per domain. To control speaker-related confounders, all utterances are synthesized using OpenAI API-based TTS \cite{hurst2024gpt} with fixed speaker assignments. We use a male target and female distractor across all distractor-language conditions. Mixtures use a default target-to-distractor SNR of 0 dB, with additional levels for robustness analysis.

\subsection{Dataset Statistics}

Target--distractor cosine similarity, measured using \texttt{multilingual-e5-large} \cite{wang2024multilingual}, ranges from 0.796 to 0.842 across distractor languages, indicating broadly comparable semantic similarity across conditions. However, non-English distractors differ in duration: Spanish and Korean distractors are approximately 3.3 s and 2.6 s longer, respectively, than the 21.4 s target average, which may affect temporal overlap. Separation quality is strongly SNR-dependent: average SI-SDR increases from 2.95 dB at $-$10 dB to 16.16 dB at $+$10 dB, with greater language-dependent variation at lower SNRs. These statistics contextualize language-conditioned performance differences beyond distractor language alone. Further details are provided in Appendix~\ref{app:dataset}. 

\begin{table*}[t]
\centering
\tiny
\setlength{\tabcolsep}{3.8pt}
\renewcommand{\arraystretch}{1.15}
\begin{tabular}{cccccccccc}
\toprule
\midrule
\multirow{2}{*}{\textbf{Group}}
& \multirow{2}{*}{\textbf{Model}}
& \multirow{2}{*}{\textbf{Setting}}
& \textbf{Target}
& \multicolumn{4}{c}{\textbf{Distractor}}
& \multirow{2}{*}{\textbf{Average}}
& \multirow{2}{*}{\textbf{Lang. Gap}} \\
\cmidrule(lr){4-4}\cmidrule(lr){5-8}
& & & \textbf{English}
& \textbf{English}
& \textbf{Spanish}
& \textbf{Korean}
& \textbf{Chinese}
& & \\
\hline
\midrule

\multirow{12}{*}{\makecell{Open-\\weight}}

& \multirow{3}{*}{\texttt{Qwen2-Audio}}
& Single
& 0.773 / 0.132 & -- & -- & -- & --
& 0.773 / 0.132 & -- \\
& & Separation
& -- & 0.482 / 0.497 & 0.477 / 0.497 & 0.707 / 0.245 & 0.450 / 0.527
& 0.529$_{(-24.4)}$ / 0.441 & 25.7 \\
& & Cocktail
& -- & 0.550 / 0.270 & 0.412 / 0.418 & 0.683 / 0.156 & 0.220 / 0.627
& 0.466$_{(-30.7)}$ / 0.366 & 46.3 \\
\cmidrule(lr){2-10}

& \multirow{3}{*}{\texttt{MERaLiON-2}}
& Single
& 0.757 / 0.176 & -- & -- & -- & --
& 0.757 / 0.176 & -- \\
& & Separation
& -- & 0.592 / 0.385 & 0.678 / 0.290 & 0.768 / 0.187 & 0.732 / 0.229
& 0.693$_{(-6.5)}$ / 0.273 & 17.7 \\
& & Cocktail
& -- & 0.500 / 0.371 & 0.608 / 0.270 & 0.655 / 0.222 & 0.642 / 0.231
& 0.601$_{(-15.6)}$ / 0.274 & 15.5 \\
\cmidrule(lr){2-10}

& \multirow{3}{*}{\texttt{Audio-Flamingo-3}}
& Single
& 0.908 / 0.030 & -- & -- & -- & --
& 0.908 / 0.030 & -- \\
& & Separation
& -- & 0.750 / 0.199 & 0.705 / 0.241 & 0.827 / 0.112 & 0.752 / 0.199
& 0.758$_{(-15.0)}$ / 0.187 & 12.2 \\
& & Cocktail
& -- & 0.560 / 0.164 & 0.440 / 0.268 & 0.753 / 0.027 & 0.567 / 0.163
& 0.580$_{(-32.8)}$ / 0.141 & 31.3 \\
\cmidrule(lr){2-10}

& \multirow{3}{*}{\texttt{Qwen2.5-Omni}}
& Single
& 0.650 / 0.061 & -- & -- & -- & --
& 0.650 / 0.061 & -- \\
& & Separation
& -- & 0.358 / 0.458 & 0.500 / 0.245 & 0.635 / 0.104 & 0.577 / 0.142
& 0.518$_{(-13.3)}$ / 0.236 & 27.7 \\
& & Cocktail
& -- & 0.250 / 0.365 & 0.300 / 0.254 & 0.547 / 0.046 & 0.307 / 0.222
& 0.351$_{(-29.9)}$ / 0.216 & 29.7 \\

\midrule

\multirow{6}{*}{\makecell{Closed-\\source}}

& \multirow{3}{*}{\texttt{GPT-4o mini Audio}}
& Single
& 0.772 / -- & -- & -- & -- & --
& 0.772 / -- & -- \\
& & Separation
& -- & 0.553 / -- & 0.577 / -- & 0.592 / -- & 0.623 / --
& 0.586$_{(-18.6)}$ / -- & 7.0 \\
& & Cocktail
& -- & 0.587 / -- & 0.593 / -- & 0.682 / -- & 0.683 / --
& 0.636$_{(-13.6)}$ / -- & 9.7 \\
\cmidrule(lr){2-10}

& \multirow{3}{*}{\texttt{Gemini-2.0-Flash}}
& Single
& 0.955 / -- & -- & -- & -- & --
& 0.955 / -- & -- \\
& & Separation
& -- & 0.892 / -- & 0.967 / -- & 0.973 / -- & 0.977 / --
& 0.952$_{(-0.3)}$ / -- & 8.5 \\
& & Cocktail
& -- & 0.677 / -- & 0.043 / -- & 0.077 / -- & 0.172 / --
& 0.242$_{(-71.3)}$ / -- & 63.3 \\

\bottomrule
\end{tabular}
\caption{Main performance across three settings at 0\,dB target-to-distractor SNR. Each cell reports Accuracy ($\uparrow$) / ECE ($\downarrow$). ECE is reported only for open-weight models with accessible token-level log-probabilities. Subscripted values in the Average column denote absolute accuracy changes, in percentage points (pp), relative to the single-stream setting. Language Gap (Lang. Gap, $\downarrow$) denotes the best--worst accuracy difference across distractor-language conditions, also in pp.}
\label{tab:main_results}
\vspace{-1em}
\end{table*}

\subsection{Diagnostic Error Taxonomy}
\label{sec:error_taxonomy}

Each item has one target-grounded correct option and three incorrect options:
\begin{itemize}[leftmargin=*, itemsep=1pt, topsep=2pt, parsep=0pt]
    \item \textbf{Target Misreasoning (Mis)}: an incorrect answer grounded in the target stream but reflecting faulty reasoning.
    \item \textbf{Distractor Interference (Int)}: an incorrect answer supported by the distractor stream.
    \item \textbf{Ungrounded Inference (Ung)}: an incorrect answer unsupported by either stream.
\end{itemize}


\section{Experimental Setup}

\subsection{Models}

We evaluate six LALMs: two closed-source models, \texttt{GPT-4o mini Audio} \cite{hurst2024gpt} and \texttt{Gemini-2.0-Flash} \cite{team2023gemini}, and four open-weight models, \texttt{Qwen2-Audio} \cite{chu2024qwen2}, \texttt{MERaLiON-2} \cite{he2024meralionaudiollmtechnicalreport}, \texttt{Audio-Flamingo-3} \cite{goel2025audio}, and \texttt{Qwen2.5-Omni} \cite{xu2025qwen2}. We follow each model's reported decoding configuration where applicable.

\subsection{Evaluation Settings and Metrics}

We evaluate three settings: single, where models receive only the single English target stream; two-stage separation, where mixtures are first separated by \texttt{ClearerVoice-Studio} \cite{zhao2025clearervoice} and then evaluated with separate streams (open-weight models answer from each stream and keep the higher-confidence response; closed-source models receive both streams in one prompt); and cocktail party, where models receive the mono (diotic) mixture \cite{you2026world} and must answer from the male English target speech while ignoring the female multilingual distractor. For SNR analysis, we additionally evaluate mixture-based settings at target-to-distractor SNRs of $-$10, $-$5, 0, +5, and +10 dB. We evaluate three fixed option permutations per question across all models, distractor languages, and SNR settings to reduce option-order bias \cite{lin2025hearing}. We report accuracy for all models and additional ECE for open-weight models with token-level log-probabilities. Decoding configuration, full prompts, separation quality, and details on metrics are provided in Appendix~\ref{app:experimental_details}.


\section{Results and Analysis}

\subsection{Overall Performance under Multilingual Interference}

Table~\ref{tab:main_results} reports the main 0\,dB results. Single-stream performance is substantially higher than cocktail party performance for every model (e.g., \texttt{Gemini-2.0-Flash}: 0.955 $\rightarrow$ 0.242), showing that single-stream understanding does not reliably transfer to source-grounded reasoning under concurrent multilingual speech. 

Separation-based inference generally improves over cocktail party inference, except for \texttt{GPT-4o mini Audio} (0.586 vs.\ 0.636), but does not restore single-stream accuracy, suggesting that reducing acoustic overlap is insufficient when source attribution remains unresolved. Indeed, at 0\,dB, open-weight models select the distractor stream for 21.5--41.0\% of items, yielding near-zero accuracy ($\leq$0.016) with high ECE ($\geq$0.73), i.e., confident wrong-stream answers (Table~\ref{tab:snr_stream_selection}). \texttt{Gemini-2.0-Flash} is a notable exception: separation nearly recovers single-stream accuracy (0.952), yet its cocktail party accuracy is the lowest among all models, suggesting a strong reliance on explicit separate streams. For open-weight models, ECE also rises sharply under interference (e.g., 0.030 $\rightarrow$ 0.141 for \texttt{Audio-Flamingo-3}), which indicates that interference degrades calibration and accuracy.

\subsection{Target-Dominance Effects across SNRs}

\begin{figure}[ht]
    \centering
    \includegraphics[width=0.9\linewidth]{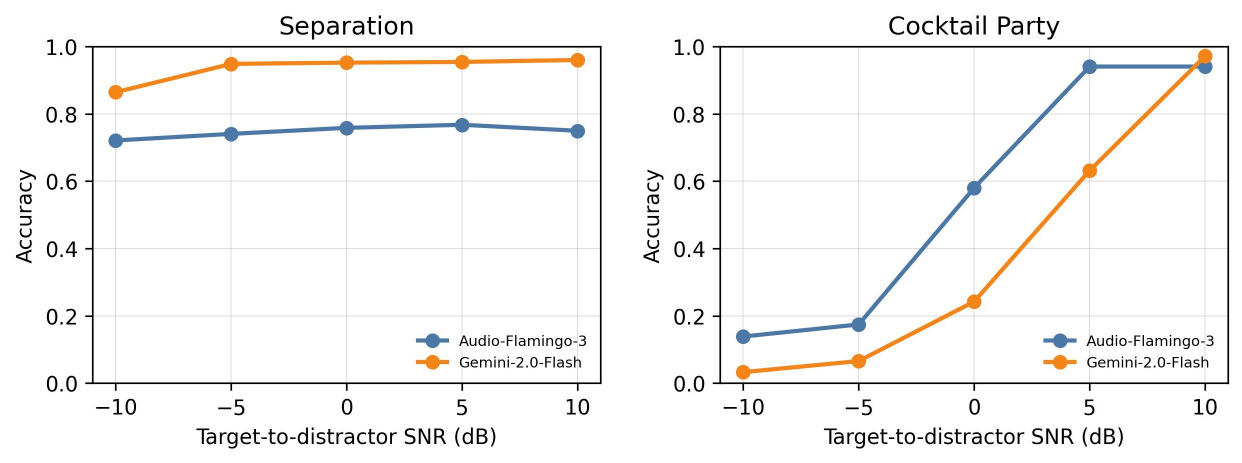}
    \caption{Average accuracy across target-to-distractor SNRs for two models under separation-based (left) and cocktail party (right) settings.}
    \label{fig:avg_acc_snrs}
    \vspace{-1.5mm}
\end{figure}

Figure~\ref{fig:avg_acc_snrs} shows that separation-based accuracy remains stable across SNRs, whereas cocktail party accuracy degrades sharply under negative SNRs and recovers as the target dominates, indicating that source grounding without separation is driven by acoustic dominance. Language gaps show a similar contrast: under cocktail party inference they widen around 0\,dB (e.g., 46.3\,pp for \texttt{Qwen2-Audio}) and fall to $\leq$6\,pp at +10\,dB for open-weight models, whereas separation-based gaps stay nearly constant (21.8--26.5\,pp for \texttt{Qwen2-Audio}; Table~\ref{tab:lang_snr_open}), suggesting model-level source preference rather than acoustic overlap. Notably, all six models exceed their single-stream accuracy at +10\,dB (up to +9.9\,pp), which we leave to future analysis.

\subsection{Evidence-Source Error Analysis}

Table~\ref{tab:error_types_cocktail} shows that the dominant failure mode shifts from target misreasoning in the single setting to distractor-grounded interference under cocktail party inference. The Int ratio reaches 0.716 for \texttt{Qwen2-Audio}, 0.807 for \texttt{Audio-Flamingo-3}, and 0.949 for \texttt{Gemini-2.0-Flash}, while Ung remains at or below 0.070. Errors therefore arise primarily from following the competing stream rather than unsupported inference.

Under cocktail party inference, Korean distractors consistently yield the lowest interference and highest accuracy for open-weight models. This is not explained by text-embedding similarity (0.815; Table~\ref{tab:dataset_statistics}) or duration, as longer Spanish distractors show no such advantage, but is consistent with Korean's higher separated-target SI-SDR at $-$10\,dB (Table~\ref{tab:separation_quality_snr}) and higher target-selection ratios (Table~\ref{tab:snr_stream_selection}). Selective-attention reliability is thus not uniform across distractor languages (Appendix~\ref{app:error_analysis}).

\begin{table}[t]
\centering
\scriptsize
\renewcommand{\arraystretch}{1.12}
\begin{tabular}{llccc}
\toprule
\midrule
\textbf{Model} & \textbf{Condition} 
& \textbf{Mis} $\downarrow$ 
& \textbf{Int} $\downarrow$ 
& \textbf{Ung} $\downarrow$ \\
\hline
\midrule

\multirow{6}{*}{\texttt{Qwen2-Audio}}
& Single             & 0.875 & --    & 0.125 \\
\cmidrule(lr){2-5}
& Cocktail Party       & 0.214 & 0.716 & 0.070 \\
& \quad w/ English     & 0.278 & 0.678 & 0.044 \\
& \quad w/ Spanish     & 0.184 & 0.725 & 0.091 \\
& \quad w/ Korean      & 0.505 & 0.347 & 0.147 \\
& \quad w/ Chinese     & 0.081 & 0.880 & 0.038 \\
\midrule

\multirow{6}{*}{\texttt{Audio-Flamingo-3}}
& Single             & 0.982 & --    & 0.018 \\
\cmidrule(lr){2-5}
& Cocktail Party       & 0.134 & 0.807 & 0.060 \\
& \quad w/ English     & 0.102 & 0.879 & 0.019 \\
& \quad w/ Spanish     & 0.095 & 0.857 & 0.048 \\
& \quad w/ Korean      & 0.324 & 0.527 & 0.149 \\
& \quad w/ Chinese     & 0.108 & 0.827 & 0.065 \\
\midrule

\multirow{6}{*}{\texttt{Gemini-2.0-Flash}}
& Single             & 0.963 & --    & 0.037 \\
\cmidrule(lr){2-5}
& Cocktail Party       & 0.040 & 0.949 & 0.011 \\
& \quad w/ English     & 0.134 & 0.830 & 0.036 \\
& \quad w/ Spanish     & 0.016 & 0.981 & 0.003 \\
& \quad w/ Korean      & 0.020 & 0.973 & 0.007 \\
& \quad w/ Chinese     & 0.052 & 0.934 & 0.014 \\
\bottomrule
\end{tabular}
\caption{Error-type distribution in the single and cocktail party (0\,dB) settings. Mis: target misreasoning; Int: distractor interference; Ung: ungrounded inference. Single has no distractor stream, so Int is not applicable.}
\label{tab:error_types_cocktail}
\vspace{-1em}
\end{table}


\section{Conclusion}

We introduce MUSI, a multilingual MCQ testbed for evaluating source-grounded LALMs under cocktail-party interference. Across six models, concurrent speech degrades performance through distractor-grounded source confusion and SNR sensitivity, with additional variation across distractor languages. These results highlight selective auditory attention as a primary capability for reliable LALMs.


\newpage

\section{Limitations}

\textbf{Scale and ecological validity.} 
MUSI comprises 200 TTS-synthesized cases (50 per domain), enabling controlled evaluation but limiting subgroup analysis. The synthetic two-speaker mixtures also omit natural conversational phenomena such as spontaneous overlap, interruptions, variable turn timing, spatial cues, channel variation, and richer prosody.

\noindent \textbf{Scope of conditions.} 
Targets are restricted to English dialogues and two-speaker mono mixtures with a single off-the-shelf separator~\cite{zhao2025clearervoice}. We use a fixed male-target/female-distractor configuration and do not counterbalance speaker gender or evaluate same-gender mixtures. This setting intentionally limits additional speaker-identity variation while focusing on multilingual semantic interference. Future work should evaluate non-English targets, bidirectional language pairs, alternative separators, and counterbalanced or same-gender speaker configurations to determine how speaker similarity interacts with selective auditory attention.

\noindent \textbf{Scope of disparity analysis.}
Our analysis concerns language-conditioned model performance, not demographic fairness, and should not be interpreted as reflecting intrinsic properties of languages or speaker communities.

\noindent \textbf{Residual confounds.}
Despite reporting length, similarity, and duration statistics (Appendix~\ref{app:dataset}), uncontrolled factors such as entity overlap, phonetic similarity, TTS voice characteristics, and temporal alignment may influence cross-lingual performance.

\noindent \textbf{Methodological asymmetry.}
Open-weight models select streams via confidence comparison, whereas closed-source models receive both streams sequentially in one prompt. This procedural difference, along with the restriction of ECE analysis to open-weight models, limits cross-group comparability.

\noindent \textbf{Evaluation format.}
The MCQ format enables controlled source attribution but does not capture free-form generation, where models may blend evidence across streams. Extending this testbed to open-ended responses remains future work.


\bibliography{custom}


\clearpage
\appendix

\onecolumn

\section{Additional Dataset Details}
\label{app:dataset}

\subsection{Construction}
We construct each item through a staged generation-and-revision pipeline: domain-specific topic and target dialogue generation, same-domain distractor generation, and diagnostic answer option construction. Text dialogues and options were produced through manual writing and LLM-assisted drafting, then manually revised. Each script is carefully streamlined to cap each stream at 30 s. All audio is sampled at 16 kHz and presented as mono (diotic) mixtures, as evaluated LALMs lack the ability to process stereo (dichotic) mixtures \cite{you2026world}.

\subsection{Quality Control}
Each item was independently validated by two annotators for target-only answerability, unique target grounding of the correct option, distractor-only support for interference options, and absence of support for ungrounded options. Annotators also checked that each question required evidence integration across target turns and did not rely on superficial lexical cues. Items were revised or removed unless all options could be traced to explicit evidence, and disagreements were resolved through author adjudication.

For Spanish, Korean, and Chinese distractors, translations were checked against the English distractor for entity preservation, event-sequence consistency, numerical consistency, and measurement-unit consistency. We used LLM-assisted back-translation as a secondary screen and manually reviewed flagged cases. We further checked for option leakage by verifying that correct, distractor-grounded, and ungrounded options were supported only by their intended evidence sources. Items with ambiguous grounding, lexical shortcuts, or cross-option leakage were revised or removed.

\subsection{Statistics}

Table~\ref{tab:dataset_statistics} reports compact dataset statistics used to assess potential length, semantic-overlap, and duration confounds across distractor languages. Script length and question length are stable by construction, while option length remains closely matched across distractor-language conditions. Target–distractor similarity, measured by \texttt{multilingual-e5-large} \cite{wang2024multilingual}, is broadly comparable across languages, although non-English distractors show slightly lower similarity than English, which may reflect cross-lingual embedding alignment as well as residual linguistic differences.

\begin{table*}[ht]
\centering
\scriptsize
\renewcommand{\arraystretch}{1.08}
\resizebox{\textwidth}{!}{%
\begin{tabular}{ccccccc}
\toprule
\midrule
\textbf{Stream}
& \textbf{Language}
& \textbf{Script Length}
& \textbf{Question Length}
& \textbf{Option Length}
& \textbf{Target--Distractor Similarity $\uparrow$}
& \textbf{Duration} \\
\hline
\midrule
Target
& English
& 45.8 $\pm$ 4.6 words
& 13.4 $\pm$ 2.2 words
& 12.1 $\pm$ 2.1 words
& --
& 21.4 $\pm$ 2.3\,s \\
\midrule
\multirow{4}{*}{Distractor}
& English
& 45.1 $\pm$ 4.9 words
& --
& 11.6 $\pm$ 2.0 words
& 0.842
& 21.7 $\pm$ 2.8\,s \\
& Spanish
& 49.6 $\pm$ 4.9 words
& --
& 11.2 $\pm$ 2.0 words
& 0.809
& 24.7 $\pm$ 2.7\,s \\
& Korean
& 158.4 $\pm$ 18.8 chars
& --
& 11.4 $\pm$ 2.0 words
& 0.815
& 24.0 $\pm$ 2.9\,s \\
& Chinese
& 81.8 $\pm$ 16.8 chars
& --
& 11.3 $\pm$ 2.0 words
& 0.796
& 22.1 $\pm$ 4.3\,s \\
\bottomrule
\end{tabular}%
}
\caption{Dataset statistics including script length, target--distractor embedding similarity, and duration. Length is measured in words for English/Spanish and characters for Korean/Chinese.}
\label{tab:dataset_statistics}
\end{table*}

\subsection{Separation Quality}

For the separation-based setting, we assess separated target quality against the corresponding single target audio using scale-invariant signal-to-distortion ratio (SI-SDR) and target automatic speech recognition word error rate (ASR WER). WER is computed between the reference target transcript and the \texttt{Whisper-v3-large} transcription of the separated stream \cite{radford2023robust}. 

Table~\ref{tab:separation_quality_snr} shows that separation quality improves with target dominance: average SI-SDR rises from 2.95 dB at $-$10 dB to 16.16 dB at $+$10 dB, while average WER drops from 0.251 to 0.066. The $-$10 dB condition shows the clearest language dependence: Korean distractors retain the best separated target quality, whereas Spanish and Chinese distractors yield substantially lower SI-SDR and higher WER. These differences narrow at 0 dB and above and indicate that language-dependent separation difficulty is most pronounced when the target is acoustically disadvantaged.

\begin{table*}[ht]
\centering
\scriptsize
\setlength{\tabcolsep}{2.5pt}
\renewcommand{\arraystretch}{1.05}
\begin{tabular}{lcccccccccc}
\toprule
\midrule
\multirow{2}{*}{\textbf{Distractor}} 
& \multicolumn{2}{c}{\textbf{-10 dB}}
& \multicolumn{2}{c}{\textbf{-5 dB}}
& \multicolumn{2}{c}{\textbf{0 dB}}
& \multicolumn{2}{c}{\textbf{+5 dB}}
& \multicolumn{2}{c}{\textbf{+10 dB}} \\
\cmidrule(lr){2-3}
\cmidrule(lr){4-5}
\cmidrule(lr){6-7}
\cmidrule(lr){8-9}
\cmidrule(lr){10-11}
& \textbf{SI} $\uparrow$ & \textbf{WER} $\downarrow$
& \textbf{SI} $\uparrow$ & \textbf{WER} $\downarrow$
& \textbf{SI} $\uparrow$ & \textbf{WER} $\downarrow$
& \textbf{SI} $\uparrow$ & \textbf{WER} $\downarrow$
& \textbf{SI} $\uparrow$ & \textbf{WER} $\downarrow$ \\
\hline
\midrule
English 
& 2.48 & 0.221
& 11.76 & 0.105
& 14.22 & 0.083
& 15.70 & 0.076
& 16.15 & 0.069 \\
Spanish 
& 1.46 & 0.372
& 12.21 & 0.180
& 14.66 & 0.148
& 15.76 & 0.121
& 16.15 & 0.088 \\
Korean 
& 7.56 & 0.147
& 12.94 & 0.072
& 14.73 & 0.062
& 15.78 & 0.058
& 16.19 & 0.051 \\
Chinese 
& 0.28 & 0.264
& 10.96 & 0.084
& 14.39 & 0.058
& 15.73 & 0.049
& 16.13 & 0.056 \\
\midrule
Average 
& 2.95 & 0.251
& 11.97 & 0.110
& 14.50 & 0.088
& 15.74 & 0.076
& 16.16 & 0.066 \\
\bottomrule
\end{tabular}
\caption{Separated target-stream quality across target-to-distractor SNRs. SI denotes SI-SDR.}
\label{tab:separation_quality_snr}
\end{table*}

\subsection{A Multilingual Dataset Example from Healthcare \& Medicine}

\noindent
{\small
\begin{examplebox}[Example Item]

\textbf{Target stream (English).}
Patient Room 402 displays symptomatic hypotension requiring an immediate
vasopressor infusion of twenty micrograms per kilogram.
Central venous pressure readings currently exceed eighteen millimeters of
mercury, indicating a significant risk of fluid volume overload.
The care team shifts to restrictive fluid management while maintaining a
baseline titration of only five micrograms.

\vspace{0.75em}
\hrule
\vspace{0.75em}

\textbf{English distractor.}
Recovery Suite B patient reports intense localized pain, prompting consideration
of an immediate ten-milligram intravenous morphine dose.
After the initial analgesic administration, respiratory monitoring detects a
decline to eight breaths per minute.
Cancel the opioid order immediately and prepare the naloxone protocol to
stabilize the patient's breathing.

\vspace{0.4em}
\textbf{Question.}
Which observed clinical indicator led to the rejection of the proposed
twenty-microgram vasopressor dosage?

\textbf{Answer options.}
\begin{enumerate}[label=(\Alph*), leftmargin=*, itemsep=1pt, topsep=2pt]
    \item Respiratory rate measurements declining to eight breaths per minute
    after analgesic exposure.
    \item Central venous pressure readings exceeding eighteen millimeters,
    indicating fluid overload risk.
    \item Reports of intense localized pain prompting opioid consideration.
    \item Electrolyte disturbances requiring counter-regulatory hormone
    administration.
\end{enumerate}

\textbf{Correct answer.} B

\textbf{Diagnostic interpretation.}
Option B is grounded in the target stream.
Options A and C are distractor-grounded, whereas Option D is unsupported by
either stream.

\vspace{0.75em}
\hrule
\vspace{0.75em}

\begin{CJK}{UTF8}{mj}
\textbf{Korean distractor.}
회복실 B의 환자가 극심한 국소 통증을 호소하여 정맥주사용 모르핀
10밀리그램의 즉시 투여가 제안되었습니다.
초기 진통제 투여 직후 호흡수가 분당 8회로 감소한 것이 감지되었습니다.
마약성 진통제 처방을 즉시 취소하고 환자의 호흡을 안정시키기 위한
나록손 프로토콜을 준비하십시오.
\end{CJK}

\vspace{0.4em}
\textbf{Question.}
Which observed clinical indicator led to the rejection of the proposed
twenty-microgram vasopressor dosage?

\textbf{Answer options.}
\begin{enumerate}[label=(\Alph*), leftmargin=*, itemsep=1pt, topsep=2pt]
    \item Reports of intense localized pain prompting opioid consideration.
    \item Respiratory rate measurements declining to eight breaths per minute
    after analgesic exposure.
    \item Central venous pressure readings exceeding eighteen millimeters,
    indicating fluid overload risk.
    \item Electrolyte disturbances requiring counter-regulatory hormone
    administration.
\end{enumerate}

\textbf{Correct answer.} C

\textbf{Diagnostic interpretation.}
Option C is grounded in the target stream.
Options A and B are distractor-grounded, whereas Option D is unsupported by
either stream.

\vspace{0.75em}
\hrule
\vspace{0.75em}

\textbf{Spanish distractor.}
El paciente de la Sala de Recuperación B refiere dolor intenso y localizado,
por lo que se propone administrar de inmediato diez miligramos de morfina por
vía intravenosa.
Tras el intento inicial de analgesia, los sensores detectan una disminución de
la frecuencia respiratoria a ocho respiraciones por minuto.
Cancele de inmediato la orden de opioides y prepare el protocolo de naloxona
para estabilizar la respiración del paciente.

\vspace{0.4em}
\textbf{Question.}
Which observed clinical indicator led to the rejection of the proposed
twenty-microgram vasopressor dosage?

\textbf{Answer options.}
\begin{enumerate}[label=(\Alph*), leftmargin=*, itemsep=1pt, topsep=2pt]
    \item Electrolyte disturbances requiring counter-regulatory hormone
    administration.
    \item Reports of intense localized pain prompting opioid consideration.
    \item Central venous pressure readings exceeding eighteen millimeters,
    indicating fluid overload risk.
    \item Respiratory rate measurements declining to eight breaths per minute
    after analgesic exposure.
\end{enumerate}

\textbf{Correct answer.} C

\textbf{Diagnostic interpretation.}
Option C is grounded in the target stream.
Options B and D are distractor-grounded, whereas Option A is unsupported by
either stream.

\vspace{0.75em}
\hrule
\vspace{0.75em}

\begin{CJK}{UTF8}{gbsn}
\textbf{Chinese distractor.}
恢复室B的患者报告剧烈局部疼痛，建议立即静脉注射10毫克吗啡。
初次尝试镇痛给药后，呼吸频率监测显示其呼吸降至每分钟8次。
请立即取消阿片类镇痛药医嘱，并准备纳洛酮处置方案，以稳定患者的
呼吸状态。
\end{CJK}

\vspace{0.4em}
\textbf{Question.}
Which observed clinical indicator led to the rejection of the proposed
twenty-microgram vasopressor dosage?

\textbf{Answer options.}
\begin{enumerate}[label=(\Alph*), leftmargin=*, itemsep=1pt, topsep=2pt]
    \item Central venous pressure readings exceeding eighteen millimeters,
    indicating fluid overload risk.
    \item Respiratory rate measurements declining to eight breaths per minute
    after analgesic exposure.
    \item Electrolyte disturbances requiring counter-regulatory hormone
    administration.
    \item Reports of intense localized pain prompting opioid consideration.
\end{enumerate}

\textbf{Correct answer.} A

\textbf{Diagnostic interpretation.}
Option A is grounded in the target stream.
Options B and D are distractor-grounded, whereas Option C is unsupported by
either stream.

\end{examplebox}
}
\normalsize


\newpage

\section{Additional Experimental Details}
\label{app:experimental_details}

\subsection{Decoding Parameters}

We follow each model's reported generation configurations, which are summarized in Table~\ref{tab:decoding_params}.

\begin{table}[ht]
\centering
\tiny
\begin{tabular}{@{}lccccc@{}}
\toprule
\midrule
\textbf{Model} & \textbf{do\_sample} & \textbf{Temp.} & \textbf{Top-$p$} & \textbf{Top-$k$} & \textbf{Max tokens} \\
\hline
\midrule
\texttt{Qwen2-Audio} & True & 0.7 & 0.5 & 20 & -- \\
\texttt{Qwen2.5-Omni} & False & 1.0 & 1.0 & 50 & -- \\
\texttt{MERaLiON-2} & -- & -- & -- & -- & 256 \\
\texttt{Audio-Flamingo-3} & True & 0.7 & 0.9 & -- & 256 \\
\bottomrule
\end{tabular}
\caption{Reported decoding parameters for open-weight LALMs. ``--'' indicates that the parameter is not explicitly specified in the model card or example used as reference.}
\label{tab:decoding_params}
\end{table}

\subsection{Source-grounding Prompts}

\noindent
{\small
\begin{promptbox}[Source-Grounding Prompt for Cocktail Party Evaluation]
You are an audio-language model being evaluated under a cocktail party test. You will receive:

\begin{itemize}[leftmargin=*, itemsep=1pt, topsep=2pt]
    \item One audio stream explicitly labeled as TARGET, spoken by a male English speaker.
    \item Zero or more audio streams explicitly labeled as DISTRACTOR.
    \item A multiple-choice question with four options.
\end{itemize}

\vspace{0.5em}
\textbf{Instructions.}
\begin{itemize}[leftmargin=*, itemsep=1pt, topsep=2pt]
    \item Base your answer exclusively on the audio labeled TARGET.
    \item Focus only on the male English speech in the TARGET audio.
    \item Completely ignore all audio labeled DISTRACTOR, regardless of language, speaker gender, or salience.
    \item Do not use information from any unlabeled or distractor audio.
    \item The correct answer requires integrating multiple pieces of evidence from the TARGET audio.
    \item Select exactly one option.
\end{itemize}

\vspace{0.5em}
\textbf{Output rules.}
\begin{itemize}[leftmargin=*, itemsep=1pt, topsep=2pt]
    \item Respond with a single capital letter only: A, B, C, or D.
    \item Do not provide explanations, reasoning steps, or additional text.
\end{itemize}
\end{promptbox}
}
\normalsize

\subsection{Separation-based Inference}
\label{app:separation}

The separation-based setting isolates acoustic overlap from source attribution: if separation restores single-level accuracy, the bottleneck is acoustic; if not, the model itself struggles to ground its reasoning in the correct stream. For open-weight models, both separated streams are processed independently, and the final response is selected from the stream with the higher normalized option confidence. This setup further diagnoses whether the model assigns greater confidence to the distractor stream than to the target stream. For closed-source models, where token-level confidence is unavailable, both separated streams are provided sequentially in one prompt, and the model must generate the answer.

\subsection{Confidence and Calibration}
For open-weight models, confidence is the softmax-normalized probability of the selected option over the log-probabilities of \{A, B, C, D\}. ECE is computed with $M=10$ equal-width bins:
\begin{equation}
\mathrm{ECE} = \sum_{m=1}^{M} \frac{|B_m|}{N}
\left| \mathrm{acc}(B_m) - \mathrm{conf}(B_m) \right|,
\end{equation}
where $N$ is the number of predictions and $\mathrm{acc}(B_m)$ and $\mathrm{conf}(B_m)$ are the accuracy and mean confidence in bin $B_m$. All ECE values, including aggregates such as the Average column in Table~\ref{tab:main_results}, are computed over predictions pooled across option permutations and conditions.


\newpage

\section{Additional Results and Analyses}
\label{app:error_analysis}

\subsection{Confidence-based Stream Selection}
Table \ref{tab:snr_stream_selection} shows that separation-based inference remains limited by source attribution. Although accuracy is high when the target stream is selected, open-weight models often assign higher confidence to the distractor stream. At 0 dB, target-selection ratios are 59.0\% for \texttt{Qwen2-Audio}, 78.5\% for \texttt{MERaLiON-2}, 78.1\% for \texttt{Audio-Flamingo-3}, and 68.9\% for \texttt{Qwen2.5-Omni}.

When the distractor stream is selected, accuracy drops sharply while ECE becomes large, indicating confident wrong-stream predictions. In particular, this misselection persists even at high SNRs, where separation quality is strong
(e.g., SI-SDR $>16$\,dB at $+10$\,dB; Table~\ref{tab:separation_quality_snr}),
suggesting that model-level source preference contributes beyond separation quality alone. Korean distractors generally yield higher target-selection ratios than the other distractor languages, consistent with the main accuracy results, although this pattern may reflect multiple factors including phonetic properties and separation quality.

\subsection{SNR-wise Error-Type Distributions}
Table \ref{tab:snr_error_types_full} shows that error types vary systematically with SNR. In the cocktail party setting, negative SNRs are dominated by distractor-grounded interference errors, indicating that models tend to follow the acoustically dominant competing stream. As the target becomes louder, the interference ratio decreases and target-misreasoning errors become more frequent. Overall ungrounded ratios remain at or below 0.14 across SNRs (up to 0.25 in individual language conditions). In the separation setting, error distributions are more stable across SNRs, but distractor-grounded errors remain common, further supporting the presence of a source-attribution bottleneck.

\subsection{Domain- and SNR-wise Performance}
Table \ref{tab:domain_snr_results} provides domain- and SNR-wise performance breakdowns. The overall trend is consistent across domains: separation-based inference remains comparatively stable (see Separation rows in Tables~\ref{tab:lang_snr_open} and~\ref{tab:lang_snr_closed}), whereas cocktail party inference is highly sensitive to target-to-distractor SNR. Negative SNRs substantially reduce accuracy, while positive SNRs improve performance as the target becomes acoustically dominant. This pattern indicates that the observed vulnerability is not restricted to a single domain, but reflects a broader limitation in source-grounded reasoning under concurrent speech. Among domains, Construction \& Plant tends to yield the lowest cocktail party accuracy at severe SNRs (e.g., 0.097 for \texttt{Audio-Flamingo-3} at $-$10 dB), possibly reflecting the confusability of its procedural and numerical
distractor content.

\subsection{SNR-wise Performance by Distractor Language}
\label{app:lang_snr}

Tables~\ref{tab:lang_snr_open} and~\ref{tab:lang_snr_closed} report accuracy by distractor language across SNRs. Under separation, language gaps stay nearly constant across SNRs (e.g., 21.8--26.5\,pp for \texttt{Qwen2-Audio}), and Korean yields the highest accuracy in 19 of 20 open-weight model--SNR cells, suggesting model-level source preference rather than residual acoustic overlap. Under cocktail party, gaps widen near acoustic parity (e.g., 46.3\,pp for \texttt{Qwen2-Audio} at 0\,dB) and shrink to $\leq$6.0\,pp at +10\,dB for all open-weight models.

\texttt{Gemini-2.0-Flash} shows a reversal. English distractors yield its highest cocktail party accuracy from $-$10 to +5\,dB (e.g., 0.955 vs.\ 0.320 for Spanish at +5\,dB) but its lowest separation accuracy at every SNR. Same-language interference thus affects stream-based and mixture-based processing differently.

\begin{table*}[ht]
\centering
\scriptsize
\renewcommand{\arraystretch}{1.06}
\resizebox{\textwidth}{!}{%
\begin{tabular}{cclccccc}
\toprule
\midrule
\multirow{2}{*}{\textbf{Model}} 
& \multirow{2}{*}{\textbf{Condition}} 
& \multirow{2}{*}{\textbf{Distractor}} 
& \multicolumn{5}{c}{\textbf{SNR}} \\
\cmidrule(lr){4-8}
& & 
& \textbf{-10 dB} 
& \textbf{-5 dB} 
& \textbf{0 dB} 
& \textbf{+5 dB} 
& \textbf{+10 dB} \\
\hline
\midrule
\multirow{11}{*}{\texttt{Qwen2-Audio}}
& Separation & Overall & 100.0\%; 0.507 / 0.445 & 100.0\%; 0.516 / 0.451 & 100.0\%; 0.529 / 0.441 & 100.0\%; 0.544 / 0.425 & 100.0\%; 0.537 / 0.429 \\
\cmidrule(lr){2-8}
& \multirow{5}{*}{Target Selected} & Overall & 59.7\%; 0.813 / 0.144 & 58.9\%; 0.867 / 0.103 & 59.0\%; 0.884 / 0.089 & 61.3\%; 0.879 / 0.094 & 60.4\%; 0.879 / 0.092 \\
&  & \quad w/  English & 54.8\%; 0.790 / 0.179 & 55.2\%; 0.870 / 0.105 & 53.8\%; 0.895 / 0.088 & 59.2\%; 0.893 / 0.094 & 56.8\%; 0.886 / 0.087 \\
&  & \quad w/  Spanish & 52.5\%; 0.787 / 0.172 & 53.3\%; 0.856 / 0.118 & 53.8\%; 0.882 / 0.098 & 55.2\%; 0.858 / 0.115 & 55.0\%; 0.882 / 0.099 \\
&  & \quad w/  Korean & 78.3\%; 0.872 / 0.088 & 76.8\%; 0.868 / 0.096 & 80.2\%; 0.863 / 0.100 & 79.0\%; 0.859 / 0.105 & 80.3\%; 0.859 / 0.105 \\
&  & \quad w/  Chinese & 53.2\%; 0.774 / 0.181 & 50.3\%; 0.874 / 0.106 & 48.3\%; 0.910 / 0.078 & 51.7\%; 0.916 / 0.069 & 49.5\%; 0.899 / 0.080 \\
\cmidrule(lr){2-8}
& \multirow{5}{*}{Distractor Selected} & Overall & 40.3\%; 0.054 / 0.896 & 41.1\%; 0.012 / 0.950 & 41.0\%; 0.016 / 0.949 & 38.8\%; 0.014 / 0.950 & 39.6\%; 0.017 / 0.946 \\
&  & \quad w/  English & 45.2\%; 0.041 / 0.913 & 44.8\%; 0.000 / 0.971 & 46.2\%; 0.000 / 0.973 & 40.8\%; 0.008 / 0.968 & 43.2\%; 0.004 / 0.965 \\
&  & \quad w/  Spanish & 47.5\%; 0.042 / 0.912 & 46.7\%; 0.000 / 0.967 & 46.2\%; 0.004 / 0.967 & 44.8\%; 0.000 / 0.973 & 45.0\%; 0.007 / 0.964 \\
&  & \quad w/  Korean & 21.7\%; 0.069 / 0.853 & 23.2\%; 0.079 / 0.825 & 19.8\%; 0.076 / 0.838 & 21.0\%; 0.063 / 0.834 & 19.7\%; 0.085 / 0.816 \\
&  & \quad w/  Chinese & 46.8\%; 0.071 / 0.884 & 49.7\%; 0.003 / 0.974 & 51.7\%; 0.019 / 0.957 & 48.3\%; 0.010 / 0.964 & 50.5\%; 0.010 / 0.964 \\
\midrule
\multirow{11}{*}{\texttt{MERaLiON-2}}
& Separation & Overall & 100.0\%; 0.683 / 0.277 & 100.0\%; 0.685 / 0.281 & 100.0\%; 0.693 / 0.273 & 100.0\%; 0.687 / 0.281 & 100.0\%; 0.688 / 0.278 \\
\cmidrule(lr){2-8}
& \multirow{5}{*}{Target Selected} & Overall & 76.7\%; 0.853 / 0.114 & 77.0\%; 0.883 / 0.088 & 78.5\%; 0.878 / 0.093 & 78.2\%; 0.875 / 0.097 & 77.8\%; 0.878 / 0.093 \\
&  & \quad w/  English & 67.7\%; 0.852 / 0.120 & 67.2\%; 0.888 / 0.098 & 67.5\%; 0.874 / 0.107 & 66.8\%; 0.888 / 0.097 & 66.3\%; 0.872 / 0.116 \\
&  & \quad w/  Spanish & 76.0\%; 0.862 / 0.104 & 73.8\%; 0.883 / 0.089 & 77.2\%; 0.875 / 0.098 & 76.0\%; 0.882 / 0.097 & 75.0\%; 0.889 / 0.084 \\
&  & \quad w/  Korean & 84.5\%; 0.858 / 0.107 & 83.8\%; 0.883 / 0.082 & 86.2\%; 0.884 / 0.084 & 86.8\%; 0.856 / 0.111 & 86.2\%; 0.857 / 0.110 \\
&  & \quad w/  Chinese & 78.7\%; 0.839 / 0.124 & 83.0\%; 0.878 / 0.093 & 83.2\%; 0.878 / 0.089 & 83.0\%; 0.880 / 0.090 & 83.5\%; 0.894 / 0.081 \\
\cmidrule(lr){2-8}
& \multirow{5}{*}{Distractor Selected} & Overall & 23.3\%; 0.125 / 0.815 & 23.0\%; 0.024 / 0.925 & 21.5\%; 0.016 / 0.929 & 21.8\%; 0.011 / 0.938 & 22.2\%; 0.022 / 0.925 \\
&  & \quad w/  English & 32.3\%; 0.082 / 0.881 & 32.8\%; 0.020 / 0.954 & 32.5\%; 0.005 / 0.963 & 33.2\%; 0.005 / 0.970 & 33.7\%; 0.010 / 0.960 \\
&  & \quad w/  Spanish & 24.0\%; 0.139 / 0.816 & 26.2\%; 0.000 / 0.953 & 22.8\%; 0.015 / 0.941 & 24.0\%; 0.000 / 0.951 & 25.0\%; 0.013 / 0.937 \\
&  & \quad w/  Korean & 15.5\%; 0.032 / 0.882 & 16.2\%; 0.031 / 0.881 & 13.8\%; 0.048 / 0.847 & 13.2\%; 0.051 / 0.878 & 13.8\%; 0.060 / 0.849 \\
&  & \quad w/  Chinese & 21.3\%; 0.242 / 0.672 & 17.0\%; 0.059 / 0.879 & 16.8\%; 0.010 / 0.916 & 17.0\%; 0.010 / 0.905 & 16.5\%; 0.030 / 0.899 \\
\midrule
\multirow{11}{*}{\texttt{Audio-Flamingo-3}}
& Separation & Overall & 100.0\%; 0.721 / 0.194 & 100.0\%; 0.740 / 0.203 & 100.0\%; 0.758 / 0.187 & 100.0\%; 0.767 / 0.180 & 100.0\%; 0.750 / 0.196 \\
\cmidrule(lr){2-8}
& \multirow{5}{*}{Target Selected} & Overall & 73.8\%; 0.938 / 0.017 & 76.4\%; 0.967 / 0.020 & 78.1\%; 0.970 / 0.018 & 79.1\%; 0.970 / 0.021 & 77.3\%; 0.969 / 0.023 \\
&  & \quad w/  English & 71.8\%; 0.947 / 0.018 & 75.0\%; 0.973 / 0.023 & 76.8\%; 0.976 / 0.024 & 77.7\%; 0.974 / 0.019 & 73.3\%; 0.968 / 0.026 \\
&  & \quad w/  Spanish & 69.7\%; 0.923 / 0.024 & 71.7\%; 0.953 / 0.025 & 73.8\%; 0.955 / 0.019 & 74.8\%; 0.958 / 0.018 & 73.8\%; 0.962 / 0.030 \\
&  & \quad w/  Korean & 84.5\%; 0.955 / 0.024 & 84.2\%; 0.956 / 0.021 & 85.3\%; 0.965 / 0.018 & 88.3\%; 0.962 / 0.015 & 86.5\%; 0.965 / 0.017 \\
&  & \quad w/  Chinese & 69.2\%; 0.925 / 0.027 & 74.8\%; 0.987 / 0.038 & 76.5\%; 0.983 / 0.027 & 75.5\%; 0.987 / 0.033 & 75.7\%; 0.982 / 0.028 \\
\cmidrule(lr){2-8}
& \multirow{5}{*}{Distractor Selected} & Overall & 26.2\%; 0.108 / 0.779 & 23.6\%; 0.005 / 0.918 & 21.9\%; 0.004 / 0.918 & 20.9\%; 0.002 / 0.923 & 22.7\%; 0.000 / 0.923 \\
&  & \quad w/  English & 28.2\%; 0.112 / 0.785 & 25.0\%; 0.007 / 0.930 & 23.2\%; 0.000 / 0.928 & 22.3\%; 0.000 / 0.932 & 26.7\%; 0.000 / 0.934 \\
&  & \quad w/  Spanish & 30.3\%; 0.093 / 0.794 & 28.3\%; 0.000 / 0.931 & 26.2\%; 0.000 / 0.934 & 25.2\%; 0.007 / 0.926 & 26.2\%; 0.000 / 0.931 \\
&  & \quad w/  Korean & 15.5\%; 0.022 / 0.871 & 15.8\%; 0.021 / 0.872 & 14.7\%; 0.023 / 0.850 & 11.7\%; 0.000 / 0.871 & 13.5\%; 0.000 / 0.884 \\
&  & \quad w/  Chinese & 30.8\%; 0.162 / 0.712 & 25.2\%; 0.000 / 0.921 & 23.5\%; 0.000 / 0.931 & 24.5\%; 0.000 / 0.938 & 24.3\%; 0.000 / 0.923 \\
\midrule
\multirow{11}{*}{\texttt{Qwen2.5-Omni}}
& Separation & Overall & 100.0\%; 0.475 / 0.238 & 100.0\%; 0.517 / 0.239 & 100.0\%; 0.517 / 0.236 & 100.0\%; 0.522 / 0.230 & 100.0\%; 0.543 / 0.211 \\
\cmidrule(lr){2-8}
& \multirow{5}{*}{Target Selected} & Overall & 65.0\%; 0.695 / 0.058 & 68.0\%; 0.748 / 0.060 & 68.9\%; 0.744 / 0.043 & 69.2\%; 0.748 / 0.066 & 71.5\%; 0.754 / 0.053 \\
&  & \quad w/  English & 44.2\%; 0.694 / 0.104 & 43.8\%; 0.768 / 0.084 & 47.2\%; 0.756 / 0.059 & 49.7\%; 0.768 / 0.076 & 50.5\%; 0.762 / 0.076 \\
&  & \quad w/  Spanish & 63.3\%; 0.697 / 0.067 & 66.2\%; 0.738 / 0.071 & 66.5\%; 0.734 / 0.046 & 68.8\%; 0.741 / 0.071 & 69.8\%; 0.733 / 0.053 \\
&  & \quad w/  Korean & 79.5\%; 0.755 / 0.050 & 81.5\%; 0.753 / 0.056 & 81.0\%; 0.778 / 0.053 & 78.8\%; 0.770 / 0.055 & 84.0\%; 0.778 / 0.047 \\
&  & \quad w/  Chinese & 72.8\%; 0.627 / 0.068 & 80.7\%; 0.740 / 0.062 & 81.0\%; 0.710 / 0.065 & 79.3\%; 0.721 / 0.074 & 81.7\%; 0.743 / 0.069 \\
\cmidrule(lr){2-8}
& \multirow{5}{*}{Distractor Selected} & Overall & 35.0\%; 0.068 / 0.636 & 32.0\%; 0.025 / 0.719 & 31.1\%; 0.016 / 0.733 & 30.8\%; 0.015 / 0.728 & 28.5\%; 0.015 / 0.728 \\
&  & \quad w/  English & 55.8\%; 0.048 / 0.741 & 56.2\%; 0.015 / 0.795 & 52.8\%; 0.003 / 0.818 & 50.3\%; 0.000 / 0.808 & 49.5\%; 0.003 / 0.803 \\
&  & \quad w/  Spanish & 36.7\%; 0.068 / 0.622 & 33.8\%; 0.044 / 0.699 & 33.5\%; 0.035 / 0.714 & 31.2\%; 0.027 / 0.712 & 30.2\%; 0.006 / 0.737 \\
&  & \quad w/  Korean & 20.5\%; 0.041 / 0.586 & 18.5\%; 0.009 / 0.655 & 19.0\%; 0.026 / 0.613 & 21.2\%; 0.016 / 0.673 & 16.0\%; 0.021 / 0.655 \\
&  & \quad w/  Chinese & 27.2\%; 0.129 / 0.478 & 19.3\%; 0.034 / 0.598 & 19.0\%; 0.009 / 0.652 & 20.7\%; 0.032 / 0.613 & 18.3\%; 0.055 / 0.572 \\
\bottomrule
\end{tabular}%
}
\caption{SNR-wise confidence-based stream selection in the Separation setting. Each SNR cell reports Selection Ratio; Accuracy / ECE, where Selection Ratio indicates the proportion of items for which the given stream was chosen.}
\label{tab:snr_stream_selection}
\end{table*}

\begin{table*}[ht]
\centering
\scriptsize
\renewcommand{\arraystretch}{1.06}
\resizebox{\textwidth}{!}{%
\begin{tabular}{cclccccc}
\toprule
\midrule
\multirow{2}{*}{\textbf{Model}} 
& \multirow{2}{*}{\textbf{Condition}} 
& \multirow{2}{*}{\textbf{Distractor}} 
& \multicolumn{5}{c}{\textbf{SNR}} \\
\cmidrule(lr){4-8}
& & 
& \textbf{-10 dB} 
& \textbf{-5 dB} 
& \textbf{0 dB} 
& \textbf{+5 dB} 
& \textbf{+10 dB} \\
\hline
\midrule
\multirow{11}{*}{\texttt{Qwen2-Audio}}
& Single & Overall & \multicolumn{5}{c}{0.875 / -- / 0.125} \\
\cmidrule(lr){2-8}
& \multirow{5}{*}{Separation} & Overall & 0.192 / 0.768 / 0.041 & 0.166 / 0.804 / 0.030 & 0.147 / 0.823 / 0.030 & 0.168 / 0.794 / 0.038 & 0.162 / 0.808 / 0.030 \\
&  & \quad w/  English & 0.182 / 0.799 / 0.018 & 0.128 / 0.862 / 0.010 & 0.100 / 0.894 / 0.006 & 0.132 / 0.865 / 0.004 & 0.114 / 0.879 / 0.007 \\
&  & \quad w/  Spanish & 0.144 / 0.832 / 0.024 & 0.147 / 0.837 / 0.015 & 0.134 / 0.850 / 0.016 & 0.155 / 0.826 / 0.019 & 0.127 / 0.850 / 0.023 \\
&  & \quad w/  Korean & 0.359 / 0.519 / 0.122 & 0.370 / 0.513 / 0.116 & 0.398 / 0.472 / 0.131 & 0.384 / 0.459 / 0.157 & 0.432 / 0.455 / 0.114 \\
&  & \quad w/  Chinese & 0.159 / 0.805 / 0.036 & 0.104 / 0.881 / 0.015 & 0.070 / 0.918 / 0.012 & 0.086 / 0.895 / 0.019 & 0.094 / 0.894 / 0.012 \\
\cmidrule(lr){2-8}
& \multirow{5}{*}{Cocktail Party} & Overall & 0.061 / 0.894 / 0.045 & 0.083 / 0.856 / 0.061 & 0.214 / 0.716 / 0.070 & 0.594 / 0.302 / 0.104 & 0.736 / 0.141 / 0.122 \\
&  & \quad w/  English & 0.023 / 0.962 / 0.016 & 0.048 / 0.929 / 0.024 & 0.278 / 0.678 / 0.044 & 0.603 / 0.348 / 0.050 & 0.748 / 0.153 / 0.099 \\
&  & \quad w/  Spanish & 0.043 / 0.931 / 0.025 & 0.055 / 0.900 / 0.046 & 0.184 / 0.725 / 0.091 & 0.650 / 0.243 / 0.107 & 0.742 / 0.113 / 0.144 \\
&  & \quad w/  Korean & 0.147 / 0.723 / 0.130 & 0.220 / 0.613 / 0.168 & 0.505 / 0.347 / 0.147 & 0.741 / 0.103 / 0.155 & 0.757 / 0.126 / 0.117 \\
&  & \quad w/  Chinese & 0.052 / 0.918 / 0.030 & 0.044 / 0.919 / 0.037 & 0.081 / 0.880 / 0.038 & 0.455 / 0.434 / 0.111 & 0.698 / 0.170 / 0.132 \\
\midrule
\multirow{11}{*}{\texttt{MERaLiON-2}}
& Single & Overall & \multicolumn{5}{c}{0.822 / -- / 0.178} \\
\cmidrule(lr){2-8}
& \multirow{5}{*}{Separation} & Overall & 0.287  /  0.657 / 0.057 & 0.235 / 0.713 / 0.052 & 0.262 / 0.696 / 0.042 & 0.279 / 0.682 / 0.039 & 0.260 / 0.705 / 0.035 \\
&  & \quad w/  English & 0.197 / 0.769 / 0.034 & 0.168 / 0.824 / 0.008 & 0.204 / 0.788 / 0.008 & 0.169 / 0.815 / 0.016 & 0.199 / 0.781 / 0.020 \\
&  & \quad w/  Spanish & 0.230 / 0.711 / 0.059 & 0.187 / 0.766 / 0.048 & 0.228 / 0.731 / 0.041 & 0.232 / 0.732 / 0.035 & 0.202 / 0.758 / 0.040 \\
&  & \quad w/  Korean & 0.389 / 0.562 / 0.049 & 0.327 / 0.562 / 0.111 & 0.360 / 0.561 / 0.079 & 0.427 / 0.513 / 0.060 & 0.414 / 0.520 / 0.066 \\
&  & \quad w/  Chinese & 0.376 / 0.532 / 0.092 & 0.312 / 0.624 / 0.064 & 0.304 / 0.634 / 0.062 & 0.366 / 0.578 / 0.056 & 0.282 / 0.698 / 0.020 \\
\cmidrule(lr){2-8}
& \multirow{5}{*}{Cocktail Party} & Overall & 0.076 / 0.858 / 0.066 & 0.108 / 0.816 / 0.075 & 0.284 / 0.587 / 0.129 & 0.602 / 0.262 / 0.136 & 0.650 / 0.243 / 0.106 \\
&  & \quad w/  English & 0.039 / 0.935 / 0.026 & 0.036 / 0.939 / 0.025 & 0.197 / 0.743 / 0.060 & 0.581 / 0.323 / 0.097 & 0.659 / 0.228 / 0.114 \\
&  & \quad w/  Spanish & 0.046 / 0.909 / 0.044 & 0.084 / 0.852 / 0.064 & 0.268 / 0.583 / 0.149 & 0.576 / 0.288 / 0.136 & 0.674 / 0.232 / 0.095 \\
&  & \quad w/  Korean & 0.118 / 0.771 / 0.111 & 0.187 / 0.694 / 0.119 & 0.357 / 0.469 / 0.174 & 0.667 / 0.179 / 0.154 & 0.658 / 0.234 / 0.108 \\
&  & \quad w/  Chinese & 0.112 / 0.793 / 0.095 & 0.149 / 0.743 / 0.108 & 0.353 / 0.488 / 0.158 & 0.588 / 0.252 / 0.160 & 0.612 / 0.282 / 0.107 \\
\midrule
\multirow{11}{*}{\texttt{Audio-Flamingo-3}}
& Single & Overall & \multicolumn{5}{c}{0.982 / -- / 0.018} \\
\cmidrule(lr){2-8}
& \multirow{5}{*}{Separation} & Overall & 0.125 / 0.837 / 0.037 & 0.101 / 0.886 / 0.013 & 0.093 / 0.891 / 0.016 & 0.108 / 0.876 / 0.016 & 0.100 / 0.882 / 0.018 \\
&  & \quad w/  English & 0.104 / 0.884 / 0.012 & 0.062 / 0.925 / 0.012 & 0.067 / 0.920 / 0.013 & 0.075 / 0.918 / 0.007 & 0.075 / 0.914 / 0.011 \\
&  & \quad w/  Spanish & 0.157 / 0.807 / 0.036 & 0.100 / 0.900 / 0.000 & 0.096 / 0.893 / 0.011 & 0.107 / 0.893 / 0.000 & 0.092 / 0.897 / 0.011 \\
&  & \quad w/  Korean & 0.202 / 0.746 / 0.053 & 0.235 / 0.722 / 0.043 & 0.183 / 0.779 / 0.038 & 0.267 / 0.678 / 0.056 & 0.232 / 0.717 / 0.051 \\
&  & \quad w/  Chinese & 0.065 / 0.882 / 0.054 & 0.045 / 0.949 / 0.006 & 0.054 / 0.940 / 0.007 & 0.046 / 0.935 / 0.020 & 0.052 / 0.935 / 0.013 \\
\cmidrule(lr){2-8}
& \multirow{5}{*}{Cocktail Party} & Overall & 0.028 / 0.937 / 0.035 & 0.034 / 0.938 / 0.027 & 0.134 / 0.807 / 0.060 & 0.860 / 0.098 / 0.042 & 0.888 / 0.035 / 0.077 \\
&  & \quad w/  English & 0.024 / 0.961 / 0.015 & 0.027 / 0.966 / 0.008 & 0.102 / 0.879 / 0.019 & 0.800 / 0.167 / 0.033 & 0.949 / 0.000 / 0.051 \\
&  & \quad w/  Spanish & 0.023 / 0.954 / 0.023 & 0.021 / 0.963 / 0.015 & 0.095 / 0.857 / 0.048 & 0.875 / 0.100 / 0.025 & 0.765 / 0.118 / 0.118 \\
&  & \quad w/  Korean & 0.061 / 0.853 / 0.086 & 0.082 / 0.847 / 0.070 & 0.324 / 0.527 / 0.149 & 0.870 / 0.043 / 0.087 & 0.878 / 0.024 / 0.098 \\
&  & \quad w/  Chinese & 0.008 / 0.970 / 0.023 & 0.016 / 0.961 / 0.023 & 0.108 / 0.827 / 0.065 & 0.889 / 0.111 / 0.000 & 0.966 / 0.000 / 0.034 \\
\midrule
\multirow{11}{*}{\texttt{Qwen2.5-Omni}}
& Single & Overall & \multicolumn{5}{c}{0.924 / -- / 0.076} \\
\cmidrule(lr){2-8}
& \multirow{5}{*}{Separation} & Overall & 0.318 / 0.610 / 0.071 & 0.321 / 0.626 / 0.053 & 0.315 / 0.634 / 0.051 & 0.310 / 0.633 / 0.058 & 0.347 / 0.598 / 0.056 \\
&  & \quad w/  English & 0.185 / 0.797 / 0.018 & 0.148 / 0.830 / 0.023 & 0.143 / 0.834 / 0.023 & 0.167 / 0.817 / 0.016 & 0.171 / 0.804 / 0.024 \\
&  & \quad w/  Spanish & 0.278 / 0.622 / 0.100 & 0.302 / 0.644 / 0.054 & 0.283 / 0.653 / 0.063 & 0.311 / 0.616 / 0.073 & 0.346 / 0.596 / 0.058 \\
&  & \quad w/  Korean & 0.409 / 0.502 / 0.089 & 0.472 / 0.433 / 0.095 & 0.447 / 0.470 / 0.082 & 0.372 / 0.534 / 0.094 & 0.461 / 0.432 / 0.107 \\
&  & \quad w/  Chinese & 0.466 / 0.436 / 0.098 & 0.483 / 0.454 / 0.063 & 0.500 / 0.449 / 0.051 & 0.458 / 0.474 / 0.067 & 0.526 / 0.417 / 0.057 \\
\cmidrule(lr){2-8}
& \multirow{5}{*}{Cocktail Party} & Overall & 0.084 / 0.843 / 0.073 & 0.111 / 0.803 / 0.086 & 0.312 / 0.589 / 0.099 & 0.721 / 0.189 / 0.090 & 0.824 / 0.091 / 0.085 \\
&  & \quad w/  English & 0.024 / 0.968 / 0.009 & 0.050 / 0.941 / 0.009 & 0.153 / 0.798 / 0.049 & 0.656 / 0.281 / 0.063 & 0.813 / 0.123 / 0.065 \\
&  & \quad w/  Spanish & 0.053 / 0.907 / 0.039 & 0.076 / 0.874 / 0.050 & 0.312 / 0.605 / 0.083 & 0.680 / 0.195 / 0.125 & 0.847 / 0.098 / 0.055 \\
&  & \quad w/  Korean & 0.185 / 0.625 / 0.189 & 0.254 / 0.493 / 0.252 & 0.621 / 0.169 / 0.210 & 0.779 / 0.093 / 0.128 & 0.811 / 0.016 / 0.173 \\
&  & \quad w/  Chinese & 0.088 / 0.843 / 0.069 & 0.095 / 0.836 / 0.069 & 0.281 / 0.623 / 0.096 & 0.790 / 0.161 / 0.048 & 0.823 / 0.114 / 0.063 \\
\midrule
\multirow{11}{*}{\texttt{GPT-4o mini Audio}}
& Single & Overall & \multicolumn{5}{c}{0.759 / -- / 0.241} \\
\cmidrule(lr){2-8}
& \multirow{5}{*}{Separation} & Overall & 0.258 / 0.612 / 0.130 & 0.265 / 0.607 / 0.128 & 0.284 / 0.595 / 0.121 & 0.280 / 0.602 / 0.118 & 0.257 / 0.642 / 0.101 \\
&  & \quad w/  English & 0.268 / 0.632 / 0.099 & 0.280 / 0.640 / 0.080 & 0.287 / 0.631 / 0.082 & 0.286 / 0.623 / 0.092 & 0.277 / 0.649 / 0.074 \\
&  & \quad w/  Spanish & 0.226 / 0.611 / 0.163 & 0.229 / 0.595 / 0.176 & 0.244 / 0.614 / 0.142 & 0.237 / 0.595 / 0.168 & 0.251 / 0.622 / 0.127 \\
&  & \quad w/  Korean & 0.278 / 0.597 / 0.125 & 0.286 / 0.607 / 0.107 & 0.335 / 0.539 / 0.127 & 0.332 / 0.574 / 0.094 & 0.252 / 0.654 / 0.094 \\
&  & \quad w/  Chinese & 0.261 / 0.606 / 0.133 & 0.265 / 0.583 / 0.152 & 0.270 / 0.593 / 0.137 & 0.273 / 0.611 / 0.116 & 0.246 / 0.642 / 0.112 \\
\cmidrule(lr){2-8}
& \multirow{5}{*}{Cocktail Party} & Overall & 0.158 / 0.767 / 0.075 & 0.227 / 0.669 / 0.104 & 0.326 / 0.552 / 0.121 & 0.432 / 0.432 / 0.136 & 0.538 / 0.333 / 0.130 \\
&  & \quad w/  English & 0.142 / 0.816 / 0.042 & 0.199 / 0.739 / 0.062 & 0.250 / 0.645 / 0.105 & 0.374 / 0.527 / 0.099 & 0.437 / 0.479 / 0.084 \\
&  & \quad w/  Spanish & 0.153 / 0.749 / 0.098 & 0.218 / 0.648 / 0.134 & 0.250 / 0.615 / 0.135 & 0.401 / 0.451 / 0.148 & 0.530 / 0.308 / 0.162 \\
&  & \quad w/  Korean & 0.161 / 0.745 / 0.093 & 0.264 / 0.639 / 0.097 & 0.440 / 0.440 / 0.120 & 0.496 / 0.359 / 0.145 & 0.681 / 0.181 / 0.138 \\
&  & \quad w/  Chinese & 0.177 / 0.756 / 0.067 & 0.233 / 0.644 / 0.122 & 0.411 / 0.463 / 0.126 & 0.509 / 0.311 / 0.179 & 0.580 / 0.260 / 0.160 \\
\midrule
\multirow{11}{*}{\texttt{Gemini-2.0-Flash}}
& Single & Overall & \multicolumn{5}{c}{0.963 / -- / 0.037} \\
\cmidrule(lr){2-8}
& \multirow{5}{*}{Separation} & Overall & 0.273 / 0.724 / 0.003 & 0.460 / 0.540 / 0.000 & 0.452 / 0.530 / 0.017 & 0.536 / 0.445 / 0.018 & 0.594 / 0.396 / 0.010 \\
&  & \quad w/  English & 0.226 / 0.774 / 0.000 & 0.129 / 0.871 / 0.000 & 0.262 / 0.738 / 0.000 & 0.226 / 0.774 / 0.000 & 0.286 / 0.714 / 0.000 \\
&  & \quad w/  Spanish & 0.354 / 0.631 / 0.015 & 0.955 / 0.045 / 0.000 & 0.750 / 0.250 / 0.000 & 0.846 / 0.077 / 0.077 & 0.952 / 0.000 / 0.048 \\
&  & \quad w/  Korean & 0.385 / 0.615 / 0.000 & 0.739 / 0.261 / 0.000 & 0.562 / 0.438 / 0.000 & 0.833 / 0.167 / 0.000 & 0.833 / 0.167 / 0.000 \\
&  & \quad w/  Chinese & 0.236 / 0.764 / 0.000 & 0.647 / 0.353 / 0.000 & 0.786 / 0.071 / 0.143 & 0.789 / 0.211 / 0.000 & 0.929 / 0.071 / 0.000 \\
\cmidrule(lr){2-8}
& \multirow{5}{*}{Cocktail Party} & Overall & 0.007 / 0.983 / 0.010 & 0.015 / 0.979 / 0.006 & 0.040 / 0.949 / 0.011 & 0.101 / 0.884 / 0.016 & 0.899 / 0.101 / 0.000 \\
&  & \quad w/  English & 0.009 / 0.988 / 0.004 & 0.028 / 0.963 / 0.010 & 0.134 / 0.830 / 0.036 & 0.741 / 0.259 / 0.000 & 1.000 / 0.000 / 0.000 \\
&  & \quad w/  Spanish & 0.005 / 0.986 / 0.008 & 0.009 / 0.988 / 0.003 & 0.016 / 0.981 / 0.003 & 0.049 / 0.944 / 0.007 & 0.760 / 0.240 / 0.000 \\
&  & \quad w/  Korean & 0.007 / 0.985 / 0.008 & 0.005 / 0.990 / 0.005 & 0.020 / 0.973 / 0.007 & 0.084 / 0.899 / 0.018 & 0.875 / 0.125 / 0.000 \\
&  & \quad w/  Chinese & 0.007 / 0.974 / 0.019 & 0.019 / 0.973 / 0.007 & 0.052 / 0.934 / 0.014 & 0.183 / 0.774 / 0.043 & 1.000 / 0.000 / 0.000 \\
\bottomrule
\end{tabular}%
}
\caption{SNR-wise error-type distribution (Mis / Int / Ung) across Single, Separation, and Cocktail Party settings. Single has no distractor stream and is therefore shown once rather than by SNR.}
\label{tab:snr_error_types_full}
\end{table*}

\begin{table*}[t]
\centering
\scriptsize
\renewcommand{\arraystretch}{1.08}
\resizebox{\textwidth}{!}{%
\begin{tabular}{cccccccc}
\toprule
\midrule
\multirow{2}{*}{\textbf{Model}} 
& \multirow{2}{*}{\textbf{Domain}} 
& \multirow{2}{*}{\textbf{Single}} 
& \multicolumn{5}{c}{\textbf{Cocktail Party (SNR)}} \\
\cmidrule(lr){4-8}
& & 
& \textbf{-10 dB} 
& \textbf{-5 dB} 
& \textbf{0 dB} 
& \textbf{+5 dB} 
& \textbf{+10 dB} \\
\hline
\midrule
\multirow{5}{*}{\texttt{Qwen2-Audio}}
& Aviation & 0.687 / 0.197 & 0.123 / 0.715 & 0.160 / 0.656 & 0.445 / 0.371 & 0.685 / 0.166 & 0.737 / 0.150 \\
& Healthcare & 0.847 / 0.090 & 0.143 / 0.730 & 0.147 / 0.707 & 0.483 / 0.375 & 0.825 / 0.093 & 0.890 / 0.061 \\
& Finance & 0.833 / 0.093 & 0.123 / 0.737 & 0.162 / 0.687 & 0.502 / 0.345 & 0.805 / 0.095 & 0.870 / 0.062 \\
& Construction & 0.727 / 0.166 & 0.092 / 0.749 & 0.138 / 0.689 & 0.435 / 0.376 & 0.708 / 0.161 & 0.795 / 0.120 \\
& Average & 0.773 / 0.132 & 0.120 / 0.731 & 0.152 / 0.685 & 0.466 / 0.366 & 0.756 / 0.124 & 0.823 / 0.093 \\
\midrule
\multirow{5}{*}{\texttt{MERaLiON-2}}
& Aviation & 0.727 / 0.191 & 0.280 / 0.513 & 0.377 / 0.423 & 0.628 / 0.241 & 0.792 / 0.119 & 0.823 / 0.098 \\
& Healthcare & 0.760 / 0.188 & 0.210 / 0.620 & 0.287 / 0.553 & 0.555 / 0.327 & 0.787 / 0.147 & 0.833 / 0.121 \\
& Finance & 0.813 / 0.143 & 0.215 / 0.615 & 0.302 / 0.520 & 0.665 / 0.215 & 0.858 / 0.076 & 0.840 / 0.102 \\
& Construction & 0.727 / 0.217 & 0.203 / 0.635 & 0.267 / 0.576 & 0.557 / 0.313 & 0.755 / 0.168 & 0.783 / 0.149 \\
& Average & 0.757 / 0.176 & 0.227 / 0.596 & 0.308 / 0.515 & 0.601 / 0.274 & 0.798 / 0.127 & 0.820 / 0.116 \\
\midrule
\multirow{5}{*}{\texttt{Audio-Flamingo-3}}
& Aviation & 0.867 / 0.038 & 0.137 / 0.635 & 0.172 / 0.571 & 0.600 / 0.092 & 0.913 / 0.032 & 0.925 / 0.023 \\
& Healthcare & 0.953 / 0.068 & 0.145 / 0.639 & 0.180 / 0.589 & 0.543 / 0.195 & 0.968 / 0.044 & 0.978 / 0.049 \\
& Finance & 0.947 / 0.043 & 0.175 / 0.590 & 0.193 / 0.552 & 0.622 / 0.121 & 0.968 / 0.036 & 0.967 / 0.031 \\
& Construction & 0.867 / 0.024 & 0.097 / 0.652 & 0.152 / 0.595 & 0.555 / 0.165 & 0.912 / 0.021 & 0.892 / 0.027 \\
& Average & 0.908 / 0.030 & 0.138 / 0.626 & 0.174 / 0.576 & 0.580 / 0.141 & 0.940 / 0.026 & 0.940 / 0.026 \\
\midrule
\multirow{5}{*}{\texttt{Qwen2.5-Omni}}
& Aviation & 0.600 / 0.123 & 0.062 / 0.544 & 0.103 / 0.467 & 0.325 / 0.213 & 0.653 / 0.064 & 0.738 / 0.055 \\
& Healthcare & 0.647 / 0.099 & 0.053 / 0.599 & 0.122 / 0.485 & 0.387 / 0.212 & 0.750 / 0.078 & 0.798 / 0.071 \\
& Finance & 0.727 / 0.048 & 0.105 / 0.499 & 0.115 / 0.459 & 0.327 / 0.233 & 0.662 / 0.027 & 0.743 / 0.046 \\
& Construction & 0.627 / 0.087 & 0.065 / 0.578 & 0.095 / 0.505 & 0.365 / 0.217 & 0.637 / 0.098 & 0.715 / 0.083 \\
& Average & 0.650 / 0.061 & 0.071 / 0.555 & 0.109 / 0.479 & 0.351 / 0.216 & 0.675 / 0.042 & 0.749 / 0.051 \\
\midrule
\multirow{5}{*}{\texttt{GPT-4o mini Audio}}
& Aviation & 0.680 / -- & 0.415 / -- & 0.527 / -- & 0.633 / -- & 0.708 / -- & 0.765 / -- \\
& Healthcare & 0.800 / -- & 0.397 / -- & 0.498 / -- & 0.675 / -- & 0.785 / -- & 0.812 / -- \\
& Finance & 0.860 / -- & 0.430 / -- & 0.527 / -- & 0.687 / -- & 0.792 / -- & 0.853 / -- \\
& Construction & 0.747 / -- & 0.365 / -- & 0.423 / -- & 0.550 / -- & 0.712 / -- & 0.773 / -- \\
& Average & 0.772 / -- & 0.402 / -- & 0.494 / -- & 0.636 / -- & 0.749 / -- & 0.801 / -- \\
\midrule
\multirow{5}{*}{\texttt{Gemini-2.0-Flash}}
& Aviation & 0.940 / -- & 0.027 / -- & 0.088 / -- & 0.292 / -- & 0.715 / -- & 0.977 / -- \\
& Healthcare & 0.960 / -- & 0.035 / -- & 0.058 / -- & 0.233 / -- & 0.608 / -- & 0.987 / -- \\
& Finance & 0.960 / -- & 0.045 / -- & 0.065 / -- & 0.238 / -- & 0.623 / -- & 0.980 / -- \\
& Construction & 0.960 / -- & 0.023 / -- & 0.050 / -- & 0.205 / -- & 0.578 / -- & 0.942 / -- \\
& Average & 0.955 / -- & 0.033 / -- & 0.065 / -- & 0.242 / -- & 0.631 / -- & 0.971 / -- \\
\bottomrule
\end{tabular}%
}
\caption{Domain-wise single and Cocktail Party performance across target-to-distractor SNRs. Each cell reports Accuracy / ECE. Single is SNR-independent and is shown once per domain.}
\label{tab:domain_snr_results}
\end{table*}

\begin{table*}[t]
\centering
\scriptsize
\setlength{\tabcolsep}{4pt}
\renewcommand{\arraystretch}{1.1}
\begin{tabular}{lllccccc}
\toprule
\midrule
\multirow{2}{*}{\textbf{Model}} & \multirow{2}{*}{\textbf{Condition}} & \multirow{2}{*}{\textbf{Distractor}} & \multicolumn{5}{c}{\textbf{SNR}} \\
\cmidrule(lr){4-8}
& & & \textbf{$-$10\,dB} & \textbf{$-$5\,dB} & \textbf{0\,dB} & \textbf{+5\,dB} & \textbf{+10\,dB} \\
\hline
\midrule
\multirow{13}{*}{\texttt{Qwen2-Audio}}
& Single & Overall & \multicolumn{5}{c}{0.773 / 0.132} \\
\cmidrule(lr){2-8}
& \multirow{6}{*}{Separation} & Overall & 0.507 / 0.445 & 0.516 / 0.451 & 0.529 / 0.441 & 0.544 / 0.425 & 0.537 / 0.429 \\
&  & \quad w/ English & 0.452 / 0.509 & 0.480 / 0.493 & 0.482 / 0.497 & 0.532 / 0.447 & 0.505 / 0.466 \\
&  & \quad w/ Spanish & 0.433 / 0.521 & 0.457 / 0.513 & 0.477 / 0.497 & 0.473 / 0.500 & 0.488 / 0.487 \\
&  & \quad w/ Korean & 0.698 / 0.251 & 0.685 / 0.263 & 0.707 / 0.245 & 0.692 / 0.258 & 0.707 / 0.244 \\
&  & \quad w/ Chinese & 0.445 / 0.508 & 0.442 / 0.534 & 0.450 / 0.527 & 0.478 / 0.497 & 0.450 / 0.525 \\
& & \quad Lang.\ Gap (pp) & 26.5 & 24.3 & 25.7 & 21.8 & 25.7 \\
\cmidrule(lr){2-8}
& \multirow{6}{*}{Cocktail Party} & Overall & 0.120 / 0.731 & 0.152 / 0.685 & 0.466 / 0.366 & 0.756 / 0.124 & 0.823 / 0.093 \\
&  & \quad w/ English & 0.038 / 0.838 & 0.090 / 0.759 & 0.550 / 0.270 & 0.765 / 0.125 & 0.815 / 0.105 \\
&  & \quad w/ Spanish & 0.077 / 0.797 & 0.087 / 0.769 & 0.412 / 0.418 & 0.767 / 0.125 & 0.838 / 0.088 \\
&  & \quad w/ Korean & 0.297 / 0.482 & 0.333 / 0.428 & 0.683 / 0.156 & 0.807 / 0.098 & 0.815 / 0.106 \\
&  & \quad w/ Chinese & 0.070 / 0.814 & 0.097 / 0.784 & 0.220 / 0.627 & 0.685 / 0.161 & 0.823 / 0.081 \\
& & \quad Lang.\ Gap (pp) & 25.8 & 24.7 & 46.3 & 12.2 & 2.3 \\
\midrule
\multirow{13}{*}{\texttt{MERaLiON-2}}
& Single & Overall & \multicolumn{5}{c}{0.757 / 0.176} \\
\cmidrule(lr){2-8}
& \multirow{6}{*}{Separation} & Overall & 0.683 / 0.277 & 0.685 / 0.281 & 0.693 / 0.273 & 0.687 / 0.281 & 0.688 / 0.278 \\
&  & \quad w/ English & 0.603 / 0.366 & 0.603 / 0.374 & 0.592 / 0.385 & 0.595 / 0.386 & 0.582 / 0.401 \\
&  & \quad w/ Spanish & 0.688 / 0.274 & 0.652 / 0.315 & 0.678 / 0.290 & 0.670 / 0.298 & 0.670 / 0.296 \\
&  & \quad w/ Korean & 0.730 / 0.227 & 0.745 / 0.210 & 0.768 / 0.187 & 0.750 / 0.212 & 0.747 / 0.207 \\
&  & \quad w/ Chinese & 0.712 / 0.241 & 0.738 / 0.227 & 0.732 / 0.229 & 0.732 / 0.229 & 0.752 / 0.214 \\
& & \quad Lang.\ Gap (pp) & 12.7 & 14.2 & 17.7 & 15.5 & 17.0 \\
\cmidrule(lr){2-8}
& \multirow{6}{*}{Cocktail Party} & Overall & 0.227 / 0.596 & 0.308 / 0.515 & 0.601 / 0.274 & 0.798 / 0.127 & 0.820 / 0.116 \\
&  & \quad w/ English & 0.152 / 0.740 & 0.213 / 0.674 & 0.500 / 0.371 & 0.793 / 0.133 & 0.795 / 0.137 \\
&  & \quad w/ Spanish & 0.175 / 0.671 & 0.267 / 0.566 & 0.608 / 0.270 & 0.792 / 0.135 & 0.842 / 0.094 \\
&  & \quad w/ Korean & 0.265 / 0.497 & 0.368 / 0.400 & 0.655 / 0.222 & 0.805 / 0.121 & 0.815 / 0.124 \\
&  & \quad w/ Chinese & 0.317 / 0.475 & 0.383 / 0.420 & 0.642 / 0.231 & 0.802 / 0.121 & 0.828 / 0.115 \\
& & \quad Lang.\ Gap (pp) & 16.5 & 17.0 & 15.5 & 1.3 & 4.7 \\
\midrule
\multirow{13}{*}{\texttt{Audio-Flamingo-3}}
& Single & Overall & \multicolumn{5}{c}{0.908 / 0.030} \\
\cmidrule(lr){2-8}
& \multirow{6}{*}{Separation} & Overall & 0.721 / 0.194 & 0.740 / 0.203 & 0.758 / 0.187 & 0.767 / 0.180 & 0.750 / 0.196 \\
&  & \quad w/ English & 0.712 / 0.211 & 0.732 / 0.218 & 0.750 / 0.199 & 0.757 / 0.195 & 0.710 / 0.239 \\
&  & \quad w/ Spanish & 0.672 / 0.236 & 0.683 / 0.258 & 0.705 / 0.241 & 0.718 / 0.227 & 0.710 / 0.235 \\
&  & \quad w/ Korean & 0.810 / 0.130 & 0.808 / 0.134 & 0.827 / 0.112 & 0.850 / 0.090 & 0.835 / 0.110 \\
&  & \quad w/ Chinese & 0.690 / 0.211 & 0.738 / 0.205 & 0.752 / 0.199 & 0.745 / 0.209 & 0.743 / 0.204 \\
& & \quad Lang.\ Gap (pp) & 13.8 & 12.5 & 12.2 & 13.2 & 12.5 \\
\cmidrule(lr){2-8}
& \multirow{6}{*}{Cocktail Party} & Overall & 0.138 / 0.626 & 0.174 / 0.576 & 0.580 / 0.141 & 0.940 / 0.026 & 0.940 / 0.026 \\
&  & \quad w/ English & 0.105 / 0.674 & 0.127 / 0.651 & 0.560 / 0.164 & 0.950 / 0.043 & 0.935 / 0.024 \\
&  & \quad w/ Spanish & 0.125 / 0.664 & 0.135 / 0.663 & 0.440 / 0.268 & 0.933 / 0.025 & 0.943 / 0.035 \\
&  & \quad w/ Korean & 0.208 / 0.498 & 0.290 / 0.393 & 0.753 / 0.027 & 0.923 / 0.027 & 0.932 / 0.017 \\
&  & \quad w/ Chinese & 0.115 / 0.676 & 0.145 / 0.608 & 0.567 / 0.163 & 0.955 / 0.036 & 0.952 / 0.042 \\
& & \quad Lang.\ Gap (pp) & 10.3 & 16.3 & 31.3 & 3.2 & 2.0 \\
\midrule
\multirow{13}{*}{\texttt{Qwen2.5-Omni}}
& Single & Overall & \multicolumn{5}{c}{0.650 / 0.061} \\
\cmidrule(lr){2-8}
& \multirow{6}{*}{Separation} & Overall & 0.475 / 0.238 & 0.517 / 0.239 & 0.517 / 0.236 & 0.522 / 0.230 & 0.543 / 0.211 \\
&  & \quad w/ English & 0.333 / 0.445 & 0.345 / 0.467 & 0.358 / 0.458 & 0.382 / 0.425 & 0.387 / 0.421 \\
&  & \quad w/ Spanish & 0.467 / 0.221 & 0.503 / 0.251 & 0.500 / 0.245 & 0.518 / 0.237 & 0.513 / 0.243 \\
&  & \quad w/ Korean & 0.608 / 0.116 & 0.615 / 0.135 & 0.635 / 0.104 & 0.610 / 0.141 & 0.657 / 0.102 \\
&  & \quad w/ Chinese & 0.492 / 0.174 & 0.603 / 0.126 & 0.577 / 0.142 & 0.578 / 0.128 & 0.617 / 0.103 \\
& & \quad Lang.\ Gap (pp) & 27.5 & 27.0 & 27.7 & 22.8 & 27.0 \\
\cmidrule(lr){2-8}
& \multirow{6}{*}{Cocktail Party} & Overall & 0.071 / 0.555 & 0.109 / 0.479 & 0.351 / 0.216 & 0.675 / 0.042 & 0.749 / 0.051 \\
&  & \quad w/ English & 0.025 / 0.738 & 0.035 / 0.684 & 0.250 / 0.365 & 0.632 / 0.082 & 0.742 / 0.093 \\
&  & \quad w/ Spanish & 0.028 / 0.632 & 0.060 / 0.543 & 0.300 / 0.254 & 0.667 / 0.054 & 0.728 / 0.049 \\
&  & \quad w/ Korean & 0.155 / 0.342 & 0.253 / 0.214 & 0.547 / 0.046 & 0.713 / 0.034 & 0.788 / 0.046 \\
&  & \quad w/ Chinese & 0.077 / 0.511 & 0.087 / 0.476 & 0.307 / 0.222 & 0.690 / 0.036 & 0.737 / 0.041 \\
& & \quad Lang.\ Gap (pp) & 13.0 & 21.8 & 29.7 & 8.2 & 6.0 \\
\bottomrule
\end{tabular}
\caption{SNR-wise accuracy by distractor language for open-weight models. Each cell reports Accuracy ($\uparrow$) / ECE ($\downarrow$). Lang.\ Gap ($\downarrow$) denotes the best--worst accuracy difference across distractor languages, in pp. Single is SNR-independent and shown once.}
\label{tab:lang_snr_open}
\end{table*}

\begin{table*}[t]
\centering
\scriptsize
\setlength{\tabcolsep}{4pt}
\renewcommand{\arraystretch}{1.1}
\begin{tabular}{lllccccc}
\toprule
\midrule
\multirow{2}{*}{\textbf{Model}} & \multirow{2}{*}{\textbf{Condition}} & \multirow{2}{*}{\textbf{Distractor}} & \multicolumn{5}{c}{\textbf{SNR}} \\
\cmidrule(lr){4-8}
& & & \textbf{$-$10\,dB} & \textbf{$-$5\,dB} & \textbf{0\,dB} & \textbf{+5\,dB} & \textbf{+10\,dB} \\
\hline
\midrule
\multirow{13}{*}{\texttt{GPT-4o mini Audio}}
& Single & Overall & \multicolumn{5}{c}{0.772 / --} \\
\cmidrule(lr){2-8}
& \multirow{6}{*}{Separation} & Overall & 0.580 / -- & 0.564 / -- & 0.586 / -- & 0.594 / -- & 0.583 / -- \\
&  & \quad w/ English & 0.547 / -- & 0.542 / -- & 0.553 / -- & 0.545 / -- & 0.525 / -- \\
&  & \quad w/ Spanish & 0.550 / -- & 0.535 / -- & 0.577 / -- & 0.563 / -- & 0.582 / -- \\
&  & \quad w/ Korean & 0.587 / -- & 0.563 / -- & 0.592 / -- & 0.628 / -- & 0.610 / -- \\
&  & \quad w/ Chinese & 0.637 / -- & 0.617 / -- & 0.623 / -- & 0.640 / -- & 0.613 / -- \\
& & \quad Lang.\ Gap (pp) & 9.0 & 8.2 & 7.0 & 9.5 & 8.8 \\
\cmidrule(lr){2-8}
& \multirow{6}{*}{Cocktail Party} & Overall & 0.402 / -- & 0.494 / -- & 0.636 / -- & 0.749 / -- & 0.801 / -- \\
&  & \quad w/ English & 0.367 / -- & 0.463 / -- & 0.587 / -- & 0.662 / -- & 0.722 / -- \\
&  & \quad w/ Spanish & 0.370 / -- & 0.442 / -- & 0.593 / -- & 0.730 / -- & 0.805 / -- \\
&  & \quad w/ Korean & 0.463 / -- & 0.520 / -- & 0.682 / -- & 0.782 / -- & 0.843 / -- \\
&  & \quad w/ Chinese & 0.407 / -- & 0.550 / -- & 0.683 / -- & 0.823 / -- & 0.833 / -- \\
& & \quad Lang.\ Gap (pp) & 9.7 & 10.8 & 9.7 & 16.2 & 12.2 \\
\midrule
\multirow{13}{*}{\texttt{Gemini-2.0-Flash}}
& Single & Overall & \multicolumn{5}{c}{0.955 / --} \\
\cmidrule(lr){2-8}
& \multirow{6}{*}{Separation} & Overall & 0.864 / -- & 0.948 / -- & 0.952 / -- & 0.954 / -- & 0.960 / -- \\
&  & \quad w/ English & 0.778 / -- & 0.897 / -- & 0.892 / -- & 0.912 / -- & 0.918 / -- \\
&  & \quad w/ Spanish & 0.892 / -- & 0.963 / -- & 0.967 / -- & 0.957 / -- & 0.965 / -- \\
&  & \quad w/ Korean & 0.935 / -- & 0.962 / -- & 0.973 / -- & 0.980 / -- & 0.980 / -- \\
&  & \quad w/ Chinese & 0.852 / -- & 0.972 / -- & 0.977 / -- & 0.968 / -- & 0.977 / -- \\
& & \quad Lang.\ Gap (pp) & 15.7 & 7.5 & 8.5 & 6.8 & 6.2 \\
\cmidrule(lr){2-8}
& \multirow{6}{*}{Cocktail Party} & Overall & 0.033 / -- & 0.065 / -- & 0.242 / -- & 0.631 / -- & 0.971 / -- \\
&  & \quad w/ English & 0.065 / -- & 0.152 / -- & 0.677 / -- & 0.955 / -- & 0.970 / -- \\
&  & \quad w/ Spanish & 0.017 / -- & 0.028 / -- & 0.043 / -- & 0.320 / -- & 0.958 / -- \\
&  & \quad w/ Korean & 0.017 / -- & 0.025 / -- & 0.077 / -- & 0.442 / -- & 0.987 / -- \\
&  & \quad w/ Chinese & 0.032 / -- & 0.057 / -- & 0.172 / -- & 0.808 / -- & 0.970 / -- \\
& & \quad Lang.\ Gap (pp) & 4.8 & 12.7 & 63.3 & 63.5 & 2.8 \\
\bottomrule
\end{tabular}
\caption{SNR-wise accuracy by distractor language for closed-source models. Separation rows provide both separated streams in a single prompt (Appendix~\ref{app:separation}). ECE is not available. Lang.\ Gap ($\downarrow$) is defined as in Table~\ref{tab:lang_snr_open}.}
\label{tab:lang_snr_closed}
\end{table*}

\end{document}